\documentclass{a_and_a/aa}  
\usepackage{multirow}
\usepackage{graphicx}
\usepackage{txfonts}
\usepackage[]{hyperref}

\hypersetup{
    colorlinks = true,
    allcolors=blue
    }

\begin{document}

   \title{Rapid early gas accretion for the inner Galactic disc}

   \subtitle{A case for a short accretion timescale}

   \author{Owain Snaith\inst{1}
         \and
         Misha Haywood\inst{1,2}
         \and
         Paola Di Matteo\inst{1,2}
         \and
         Matthew Lehnert\inst{2}\thanks{Current address:  Université Lyon 1, ENS de Lyon, CNRS UMR5574, Centre de Recherche Astrophysique de Lyon, F-69230 Saint-Genis-Laval, France}
         \and
         David Katz\inst{1}
         \and
         Sergey Khoperskov\inst{3, 4,1}
          }

   \institute{GEPI, Observatoire de Paris, PSL Research University, CNRS, Place Jules Janssen, 92190, Meudon, France
         \and
             Sorbonne Université, CNRS UMR 7095, Institut d'Astrophysique de Paris, 98bis bd Arago, 75014, Paris, France
        \and
        Leibniz-Institut für Astrophysik Potsdam (AIP), An der Sternwarte 16, 14482 Potsdam, Germany
        \and
       Institute of Astronomy, Russian Academy of Sciences, 48 Pyatnitskya St., Moscow 119017, Russia
             }

   \date{\today}

 
  \abstract
  {Recent observations of the Milky Way and galaxies at high redshifts suggest that galaxy discs were already in place soon after the Big Bang. While the gas infall history of the Milky Way in the inner disc has long been assumed to be characterised by a short accretion time scale, this has not been directly constrained using observations.  
  }
  {Using the unprecedented amount and quality of data of the inner regions of the Milky Way that has recently been produced by APOGEE and Gaia, we aim to derive strong constraints on the infall history of the inner ($<$ 6~kpc) Galaxy  (with a focus on stars between 4-6 kpc, which we show is an appropriate proxy for the entire inner disc).}
   {We have implemented gas infall into a chemical evolution model of the Galaxy disc, and used a Schmidt- Kennicutt law to connect the infall to the star formation. We explore a number of models, and two different formulations of the infall law. In one formulation, the infall is non-parametric, and in the other the infall has an explicitly exponential form. We fit the model parameters to the time-[Si/Fe] distribution of solar vicinity stars, and the metallicity and [Si/Fe] distribution function of stars with a galactocentric radius between 4–6 kpc from APOGEE. 
}
   {Our results point to a fast early gas accretion, with an upper limit of accretion timescale of around 2 Gyr in the inner disc of the Milky Way. This suggests that at least half the baryons were in place within 2--3 Gyr of the Big Bang, and that half the stars  of the inner disc formed within the first 5 Gyr, during the thick disc formation phase. This implies that the stellar mass of the inner disc is dominated by the thick disc, supporting our previous work, and that the gas accretion onto the inner disc was rapid and early. 
   }
{}


   \maketitle

\section{Introduction}

Understanding the history of the Milky Way requires a combination of observations   \citep[e.g.][]{Ahumada2019, Hayden2015, Adibekian2012, Queiroz2020}, simulations  \citep[e.g.][]{Guedes2011, Brook2012, Roskar2013, Brook2020, Buck2020}, models  \citep[e.g.][]{Chiappini1997, Cote2017, Spitoni2019} and statistical analysis  \citep{Ness2019,Ciuca2020}, because each approach offers different insights into the formation of our galaxy. The long timescale for many astrophysical processes means that we usually cannot see them taking place directly, and so must rely on models and simulations, which can then be compared to observations. 

 Recent surveys, such as Gaia \citep{Prusti2016, Brown2018}, APOGEE \citep{Nidever2014, Majewski2017},GALAH \citep{Buder2018}, LAMOST \citep{Cui2012}, and local spectroscopic surveys such as the one carried out by \citet{Adibekian2012}, have provided detailed information about the properties of stars. Recently, accurate stellar ages \citep[e.g.][]{Haywood2013, Silva2018, Mackereth2019} have allowed us to constrain the star formation history (SFH) of the Galaxy \citep{Snaith2014,Snaith2015,Spitoni2019}, to a degree which was previously impossible. This has resulted in an evolution in our understanding of the transition from early  to late star formation, and brought the timescale for the onset of the build up of the thick/thin disc forward from very early times \citep[e.g.][]{Chiappini1997} to later times \citep[e.g.][]{Snaith2014,Spitoni2019,Spitoni2020}. 
 
Since we presented our model of the star formation history of the Milky Way in \citet{Snaith2014,Snaith2015}  new observations from APOGEE \citep{Nidever2014, Ahumada2019, Hayden2015} have been released, and have been studied in detail. These developments mean that our chemical evolution model must be compared to these newer data sets. Since its first publication, however, our `closed-box' model has remained highly robust to comparisons with data \citep[e.g.][]{Haywood2015, Haywood2016a, Haywood2016, Haywood2018, Haywood2019}. 

Solar vicinity data from \citet{Adibekian2012}, with precise ages from \citet{Haywood2013}, offer insight into the makeup of the Milky Way. Many studies, such as \citet{Chiappini1997}, up to more recent studies, such as \citet{Grisoni2018} and \citet{Spitoni2019}, attempt to fit all the stars in the solar vicinity with a single chemical evolution. This evolution is tuned to account for both the high and low $\alpha$ sequence in the [Fe/H]-[$\alpha$/Fe] distribution. The picture is made more complicated, because stars do not remain at their place of origin, but move through the disc due to the action of various processes  \citep[e.g.][]{Sellwood2002, SanchBla2009, Schonrich2009, Minchev2011, RadburnSmith2012, Kubryk2015, Halle2018}. Thus, any given region of the disc incorporates stars that evolved at wide range of birth radii. This means they can potentially have formed in different environments, with different evolutions.  The high $\alpha$ sequence is, then, associated with stars which formed in the inner region, which we  considered to be essentially 'closed box', after a period of rapid early infall. However, this was considered to be an approximation, because it is known from both observations and cosmological simulations that a degree of infall continues until the present day, despite the infall rate falling off with time \citep[e.g.][]{Dekel2009}. 

Other authors have found Milky Way-like parallel sequences in the [$\alpha$/Fe]-[Fe/H] relation in simulations. For example simulations by \citet{Brook2004} and \citet{Brook2005} have associated the dual sequences with high early star formation caused by mergers at early times.  The simulations of \citet{Buck2020} also produce a low-alpha sequence through mergers, while, more recently, \citet{Renaud2021} have shown that $\alpha$-enhanced stars may have formed during starburst episodes associated with mergers, while the low-alpha sequence may be the result of an in-situ, more quiescent star formation.  In these examples, interactions with other objects are used to build up the two sequences rather than purely stochastic processes. In our modelling we are less interested in the origin of our accreted gas as much its impact on the chemical evolution and overall accretion rate.

From spectroscopic survey data we can see that there is a split in the evolution of the [Fe/H]--[$\alpha$/Fe] distribution into two sequences according to the distance from the galactic centre\citep{Bensby2011, Bovy2012, Haywood2013, Ahumada2019, Leung2019, Queiroz2020}. The outer disc corresponds to the low alpha sequence, and may require dilution at later times in order to mimic observations \citep{Snaith2015}, while the inner regions correspond to the distinct high alpha sequence.

APOGEE shows that the low alpha sequence is increasingly dominant with increasing radius \citep{Hayden2015, Leung2019, Queiroz2020, Anders2014}. This implies that stars in the high and low alpha sequences in the solar vicinity have different origins, meaning that each sequence should be fitted using its own separate evolution. This is re-enforced by results from dynamical simulations \citep[e.g.][]{Halle2015, Halle2018, Khoperskov2020}, which show that the Outer Lindblad Resonance (OLR) of the bar can suppress migration across it, although the OLR region itself is found to be quite broad.This effectively makes the regions inside and outside the OLR region separate environments by suppressing mixing. This  may  allow the inner and outer discs to have a different evolution, manifesting in the two sequences in [Si/Fe]-[Fe/H].  Thus,  the traditional galactic chemical evolution approach for the solar vicinity \citep{Chiappini1997,Spitoni2019}, which tries to fit both the high and low $\alpha$ sequences to a single evolution, is not the only way, or necessarily the best way, to approach modelling the solar vicinity.

There is increasing evidence that the thick disc is a massive component of the Milky Way \citep{Haywood2013,Snaith2014, Snaith2015} and that galaxy discs can build up rapidly at high redshift \citep{Neeleman2020}. For example, very low metallicity stars (with [Fe/H]<-4) have been found on disc orbits, suggesting that disc assembly can begin very early \citep{Sestito2019, Sestito2020, DiMatteo2020}. A detailed discussion of this scenario for the structure of the Milky Way can be found in \citet{Haywood2019}, including an explanation of how infall in the outer regions of the disc produces the low-alpha sequence. \citet{Haywood2019} presented an analysis which shows that the Sun is an 'outer disc' star, with properties characteristic of the low $\alpha$ sequence. The distribution of stars in abundance-metallicity space is thought to be built up through contributions from separate regions of the disc with distinct evolutions, as described in \citet{Snaith2015}. The separate evolutions for the inner  and outer discs are discussed in detail in \citet{Haywood2018} and \citet{Haywood2019} respectively. 

 There are a number of approximations used in our chemical evolution model \citep{Snaith2014}, which requires testing in detail. For the inner disc, we assumed that metals can be diluted by all the gas which is eligible to form stars across cosmic time, i.e. a very large reservoir is in place from the beginning of the model's run. This is implemented with the assumption that the total amount of gas present at $t=0$ is equal to unity, and the integral of the star formation history is also equal to one. The total stellar mass in these normalised units is around $0.7$ because of gas released back into the ISM due to stellar evolution \citep[and depends on the IMF,][]{Snaith2015}.   

We have argued, however, that our model is a first order approximation of where the majority of infall to the inner disc takes place early. The exact shape of the star formation history is then assumed to be sculpted by other processes on top of the Schmidt-Kennicutt law \citep{Kennicutt1998} normally used in galactic chemical evolution models to convert gas into stars. 

This is in contradiction to the classical understanding of galaxy evolution, where the infall of gas is important for shaping galaxies \citep[e.g.][]{Dekel2006}. In this paper we seek to open the box and allow explicit infall onto the galaxy in order to resolve the differences between our model with no infall and the results of observations, simulations and models that provide evidence of explicit infall. 

This paper will discuss the data we will use to tune our model in Section \ref{Sec:Data}, we will go on to briefly outline our chemical evolution model in Section \ref{Sec:Model}. In Section \ref{Sec:MainResults} we will  explore various scenarios and parameter combinations using our model, and then present the key results from our preferred scenario. We will discuss and contextualise our results in Section \ref{Sec:Discuss}, and, finally, present our conclusions in Section \ref{Sec:Conc}. 

\section{Data}
\label{Sec:Data}

\begin{figure*} 
\centering
	\includegraphics[scale=.6]{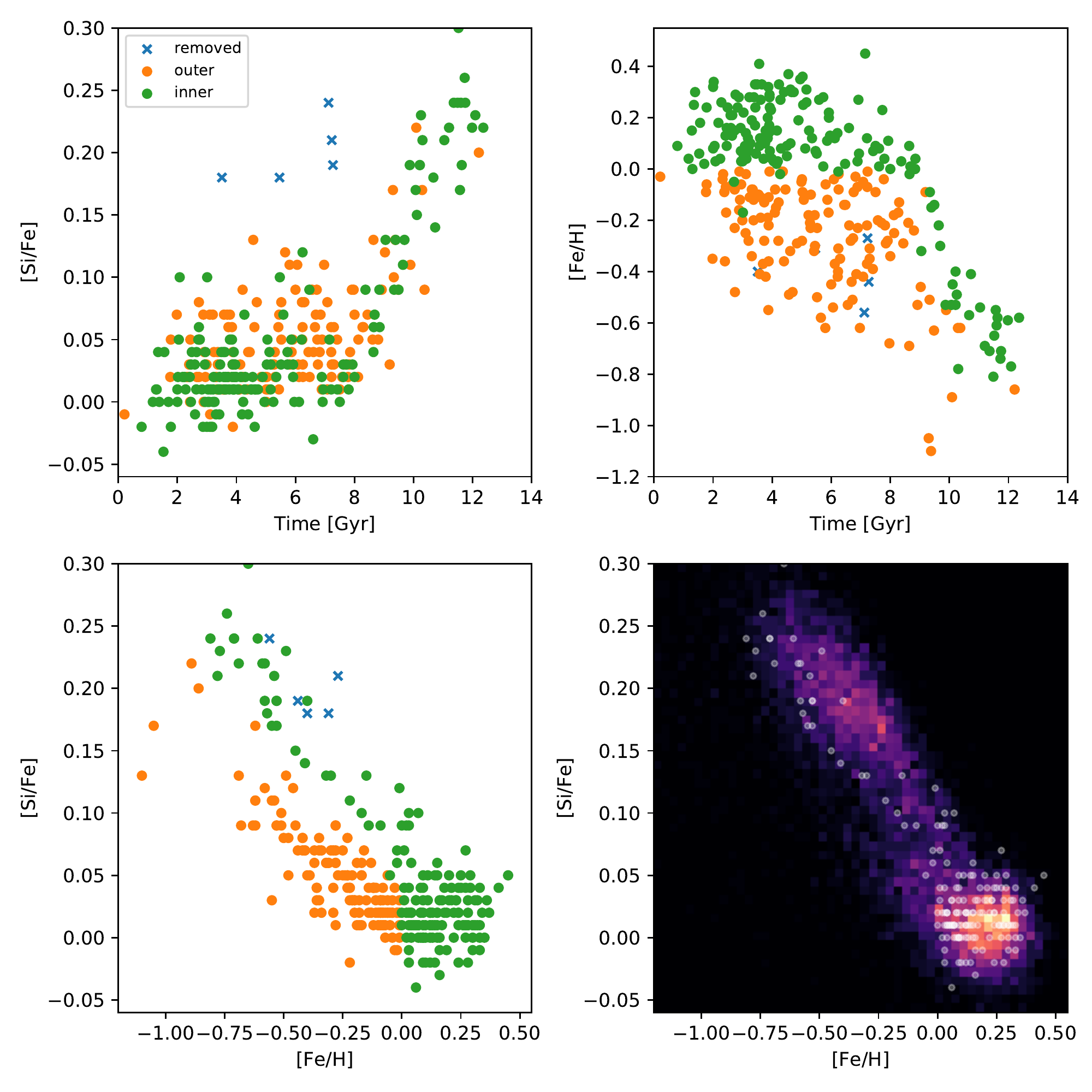}
	\caption{Top row and bottom left panel: Decomposition of the chemistry-age distribution of stars from \citet{DelgadoMena2017} with ages by Haywood et al. in prep. In the top row and bottom left panels the stars are coloured by whether it is part of the inner or outer disc. Several young alpha rich stars, \citep{Haywood2013} shown in blue, are removed from the sample. The bottom right panel shows the distribution of inner disc stars taken from APOGEE, with the inner disc stars from \citet{Adibekian2012,DelgadoMena2017} plotted as white points.  }
	\label{fig:obsdata}
\end{figure*}

We make use of data from \citet{DelgadoMena2017}, which has highly precise spectroscopy of 1111 FGK stars from \citet{Adibekian2012}, and combine this with Gaia DR2 parallaxes . These data (with the exception of Gaia) were also used by \citet{Haywood2013} to derive the ages of stars using a Bayesian estimate for each star. The entire \citet{DelgadoMena2017} sample was pruned so that the age sample contained only robust estimates. These ages have a formal error of $<$1 Gyr random and 1 Gyr systematic \citep[i.e. there is a systematic change in the calculated ages if different stellar isochrones are used,][]{Haywood2013}. These data provide us with elemental abundance ratios for a range of elements including [Fe/H], [Mg/H] and [Si/H], as well as stellar ages for a total of 301 stars, after pruning.

From the chemistry of solar vicinity stars we can separate the stellar populations between the inner and outer discs \citep[e.g.][]{Haywood2019}. This  division is justified by both observations \citep{Queiroz2020} and dynamical simulations \citep{Buck2020}. In addition to the split between inner and outer disc stars, the inner disc itself can be decomposed into the thick disc and an inner thin disc. Similarly to \citet{Snaith2015} and \citet{Haywood2013}, the outer disc is defined according to the low alpha sequence as those stars which obey:

\begin{equation}
        \begin{aligned}
\mbox{[Si/Fe]} &< -\mbox{[Fe/H]}/6. + 1./15. \\
\mbox{[Fe/H]}&<-0.0     
        \end{aligned}
\label{Eqn:Outerdisc}
\end{equation} 
\noindent
Our inner disc sample contains a total of 161 stars in the abundance-time distribution, with the remainder being assigned to the outer disc or discarded (see Fig. \ref{fig:obsdata}).  Of the remaining stars, the thick disc and inner thin disc are separated by the knee of the age-[Si/Fe] distribution at around 8 Gyr ago. This is a logical place to subdivide the sample into different populations, because it marks a strong change in the behaviour of the chemical evolution (specifically the [Si/Fe], but this can also be seen in other elements such as Mg) of the stars. 

In addition to this solar vicinity data we also make use of data from APOGEE \citep[e.g.][]{Nidever2014, Majewski2017}, specifically from SDSS DR16 \citep{Ahumada2019} with distances taken from the \texttt{Starhorse} catalogue by \citet{Queiroz2020}. We select data where $4<R_{gal}<6$~kpc, where $R_{gal}$ is the current distance of the star from the centre of the galaxy. This sample of 8871  stars  is then used to calculate the metallicity and abundance distribution functions (MDF and $\alpha$DF respectively). This provides us with a signature of the chemical evolution in the inner disc which samples the number of stars with a given metallicity and abundance. This encodes additional information, compared to the time-abundance distribution, which provides the locus of the regions of age-chemistry space sampled by stars. The  distribution of stars we use to constrain the MDF and $\alpha$DF of the inner disc are shown in the bottom right hand panel of Figure \ref{fig:obsdata}. We use the [Fe/H] and [Si/Fe] values drawn from the \texttt{astronn} VAC \citep{Leung2019}.We do not, however, use stellar ages from APOGEE, instead relaying on the ages and abundances from \citet{DelgadoMena2017}.

We seek to capture the {\it average} properties of stars in the inner disc using our one-zone model. Other works \citep[e.g.][]{Spitoni2019} make use of stars in the solar vicinity (approximately 8 kpc from the centre of the Galaxy), but, because the solar vicinity appears to be a transition region between the inner and outer disc  \citep[separated by the OLR region, with a width of several kpc at around 9--10 kpc,][]{Khoperskov2020}, stars close to the Sun are not representative of just the inner disc.  Instead, we limit our range to less than 6 kpc to avoid contamination by outer disc stars, and greater than 4 kpc to diminish contamination by the bulge and bar. 
 We take the stars in this region to be representative of the entire inner disc. This will be discussed in greater detail in Section \ref{Sec:radiusprof}.

The outer disc and young alpha rich stars \citep{Haywood2013} are removed from the sample, and we will focus on the inner disc only. We also show a comparison between the stars in the  \citet{DelgadoMena2017} sample and the stars selected from APOGEE in the bottom right panel of Fig. \ref{fig:obsdata}. One difference is that the  \citet{DelgadoMena2017} data seems to be representative of the lower limit of the [Si/Fe]--[Fe/H] between [Fe/H]  = -0.75 and [Fe/H]  = 0. We also see two density peaks in the distribution, also seen in \citet{Queiroz2020}, along the evolution of the upper sequence. It is interesting to note that the lower metallicity peak is very poorly sampled by our subsample of \citet{DelgadoMena2017} with good ages. This may be due to the fact that the \citet{DelgadoMena2017} data do not extend far from the plane, while the velocity dispersion in the z-direction is large at lower metallicities. The fact that the \citet{DelgadoMena2017}  data samples the lower edge of the APOGEE thick disc may be due to a small offset in the two catalogues, or due to the relatively small numbers of stars in that region of the plot. 

\subsection{Selection Function}
\label{Sec:SF}

 For careful comparison between the chemical evolution model and the MDF and $\alpha$DF of APOGEE we must consider the selection function of the survey. Following the method of \citet{Fragkoudi2018}, we make use of an N-body simulation containing $1 \times 10^7$ particles, and composed of three discs with different chemical distributions. The stars assigned to each of the three discs (thin disc, intermediate disc and thick disc) are taken randomly from under a Gaussian distribution in [Fe/H] and [$\alpha$/Fe], with a given centre and width. The chemical and morphological properties of these discs are given in Table 1 of \citet{Fragkoudi2018}.\footnote{$[\alpha/Fe]$ dex for D1 should be approximately 0.01 in order to reproduce the distribution shown in their Figure 3.} This approach will allow us to explore the importance of the selection function, both in terms of the impact of the different pointings on the sky, and the distance effects due to the magnitude limited nature of the survey.

We centre the simulated galaxy at the density peak of the disc, and place the Sun such that it lies at x = -8 kpc. We rescale the simulation such that the bar has a length of 4.5 kpc in order to match the Milky Way bar length found in \citet{Bland2016}. 

We assume that the inner disc has a flat metallicity gradient, which matches well with our one-zone chemical evolution model. We seek to identify the average SFH and chemical evolution of the inner region (or at least the region between 4--6 kpc) and so both the N-body model and GCE model are consistent in their assumptions. For more details about the simulation please see section 2 of \citet{Fragkoudi2018},  with further discussion in \citet{Khoperskov2020A}.

We recover the direction and radius of each  APOGEE pointing from the survey. In order to select the stars in the N-body model along each line of sight we convert the model into galactic coordinates using \texttt{Astropy} \citep{astropy2013, astropy2018}. We then select all stars in the simulation along each pointing using \texttt{Balltree} from \texttt{sci-kit learn} \citep{scikit-learn}. Thus, for each APOGEE pointing we have a corresponding field through the N-body simulation.

\begin{figure*} 
  \centering
 \includegraphics[scale=.50]{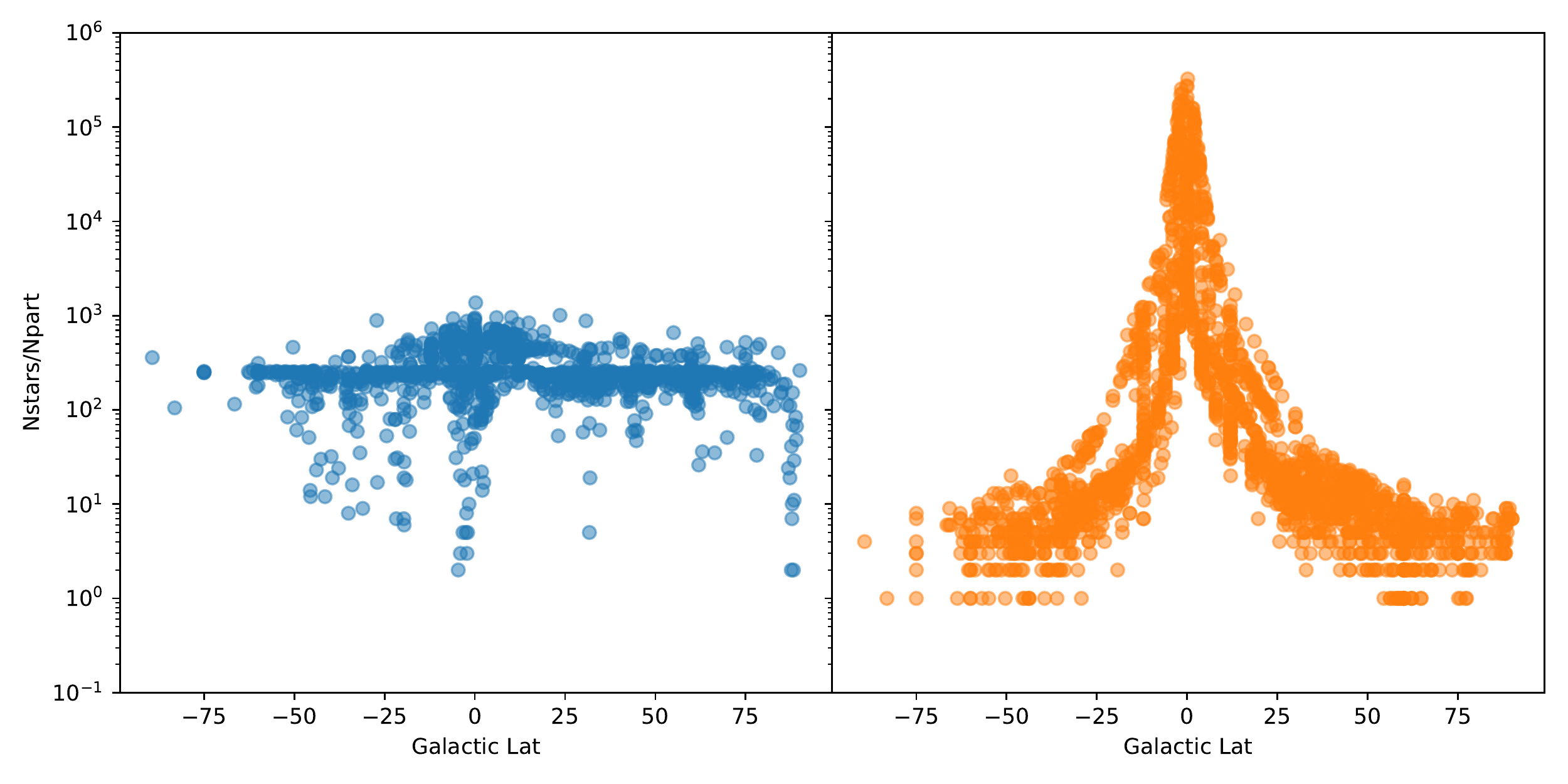}
  \caption{Left: The number of stars in each APOGEE pointing from the APOGEE survey varying over galactic latitude. Right: The number of star particles in each APOGEE pointing from the N-body simulation. }
  \label{fig:nstarpart}
\end{figure*}

The N-body simulation produces far fewer star particles in the fields at high galactic latitude and more at low galactic latitude relative to the survey, and this is depicted in Figure \ref{fig:nstarpart}. Further, each star particle in the simulation consists of approximately 8$\times$10$^3$ individual stars (assuming they have 1 solar mass). The low number of star particles at high latitude means that the stellar density field is poorly sampled, making it more difficult to study the selection function. In order to mitigate this effect we re-bin the APOGEE pointings in order to find the selection function of a collection of pointings. We use a varying number of bins with galactic latitude, following the gridding procedure  from \citet{Duarte2014}. The number of bins varies over latitude according to,

\begin{equation}
B_{l} = cos[b_{min} + (\delta b  (i_{lon} + .5))]\times 20 \times \frac{n_{lon}}{n_{lat} },
\end{equation}
\noindent where $\delta b$ is the range in latitude over the number of latitude bins. We set the number of latitude bins to 15 and 20 longitude bins at the equator. 

In each bin we identify the distance distribution from the APOGEE survey by generating a histogram of the stars, binned by distance, between 0 and 20 kpc. For each pointing we use these histograms to generate a cumulative probability function. We then  assign each star from the N-body simulation a weighting by interpolating along this probability function  and generate a random number between 0 and 1. We select the first star particle, ordered by distance, where its weighting exceeds the random number. We repeat this process, {\it with replacement}, until we have the same number of star particles as there are in the APOGEE survey pointings or bins. 

\begin{figure*} 
  \centering
 \includegraphics[scale=.50]{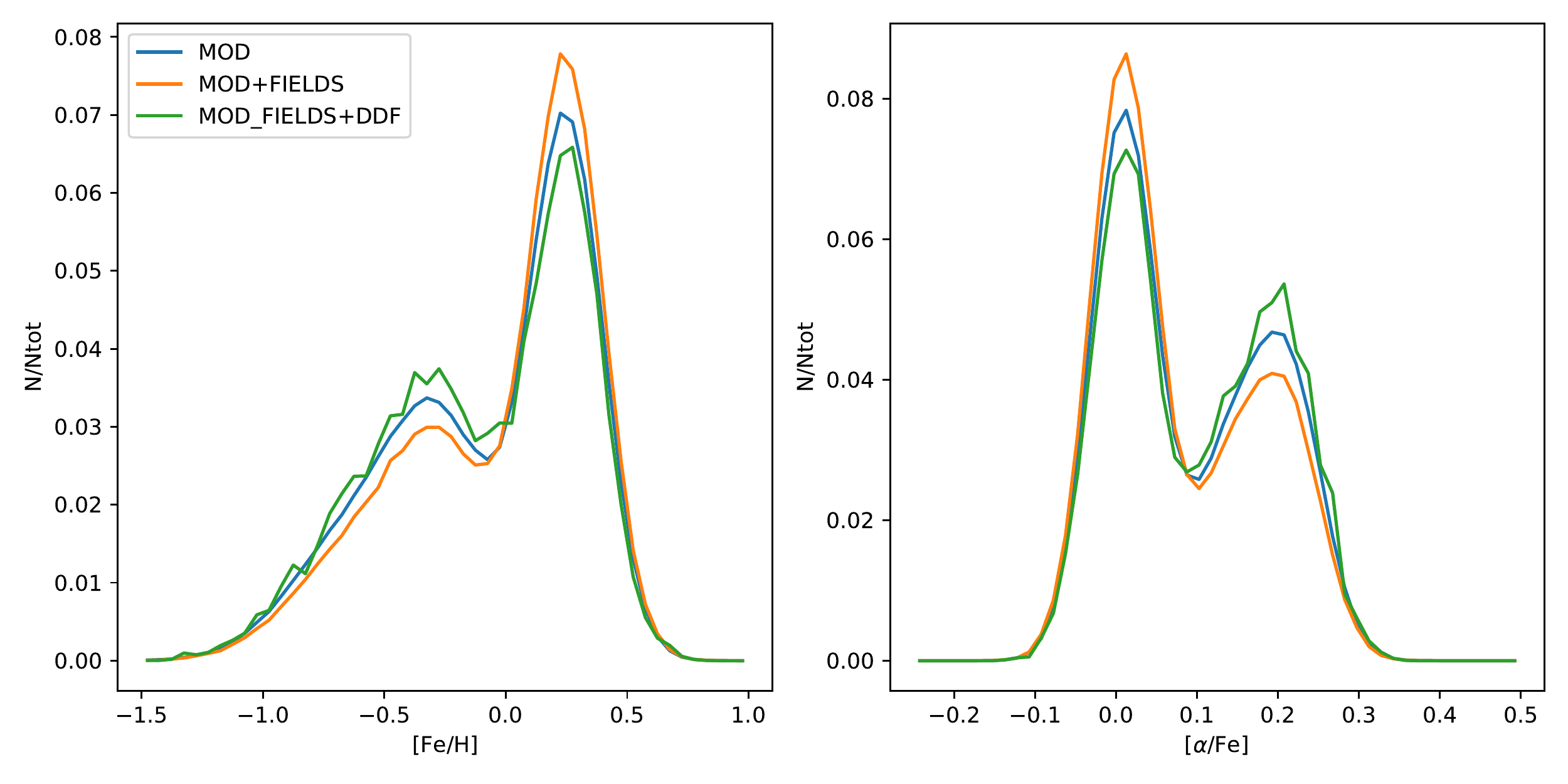}
  \caption{Left panel: The MDF, right: The $\alpha$DF. The blue line is the DFs of all star particles in the N-body model, the orange line shows the DFs of all particles along each APOGEE pointing, and the green line shows the DFs of the star particles along each pointing, with the same distance distribution function (DDF) as APOGEE. We show the MDF and $\alpha$DF for stars with a galactocentric radius of 4-6 kpc.}
  \label{fig:distributionNbody}
\end{figure*}

There are differences between the overall N-body simulation MDF and the reconstructed MDF for individual pointings \citep{Fragkoudi2018}, but this seems to be less true as we sum all the various pointings because the model and reconstructed MDFs strongly converge. 

In Figure \ref{fig:distributionNbody} we show the effect of the above selections on the N-body MDF and $\alpha$DF between 4--6 kpc. The blue line shows the original distribution functions of the simulation, the orange line shows the distribution functions for all star particles along each APOGEE pointing/bin, and the green lines shows the result of the entire process, including matching the distance distribution function (DDF). The relatively small deviation of the final DFs from the original DFs suggests that convolving the APOGEE data with a complex selection function is unnecessary for our purposes. Given the uncertainties in comparing the observations, the simulation, the expected true distribution of stars, and the relatively small deviation between the different distributions in Fig. \ref{fig:distributionNbody}, we chose to use the MDF and $\alpha$DF from the APOGEE sample without deconvolution. 

There are additional factors in the selection function which may influence the shape of the distribution functions. For example,  we have not fully accounted for the three magnitude cohorts in APOGEE, where different pointings can reach different magnitude limits. The processing of APOGEE can give a greater weighting to stars in the brighter cohorts. This would impact our sample because there is a luminosity difference between metal rich and metal poor stars. This is because, at a given mass, metal poor stars have a higher effective temperature and luminosity. Furthermore, thin disc stars can be expected to suffer from higher extinction than thick disc stars, because extinction is higher at lower Galactic latitudes. \citet{Majewski2017}  shows that extinction can cause the thick disc population to extend to greater distances than the thin disc stars. This can, therefore, be expected to impact the shape of the MDF, biasing it towards thick disc stars. If this effect is considerable, the shape of the MDF and $\alpha$DF can change, pushing infall and star formation to a different redshift. Thus, certain selection effects may remain in our data. A detailed calculation of the distribution functions is beyond the scope of this paper, but it must be noted that the results presented here, and the effects of the selection function, may not have been fully disentangled. For this paper, however, we assume these to be second order effects. 

An additional factor that may affect the results is the impact of radial migration. As there is no dynamical information in a chemical evolution model we cannot hope to replicate the impact of churning \citep{Sellwood2002}. 

The \texttt{astronn} VAC from APOGEE gives us access to a guiding radius for the orbits. This has the advantage of reducing the effect of blurring, but requires various approximations in terms of the shape of the Milky Way potential.  We discuss this in detail in Section \ref{sec:guiding}

\section{Models}
\label{Sec:Model}

The model discussed in this paper makes use of the chemical yields and overall design of the model presented in \citet{Snaith2014,Snaith2015}. 

Our original  model calculated the chemical evolution of the solar vicinity, using a sample of stars similar to the one described in the previous section (and shown in the top row and bottom left panels of Fig. \ref{fig:obsdata}). The \citet{Snaith2015} chemical evolution model differs from the `traditional' type of model in two ways, firstly, it is designed to fit the high and low alpha sequences separately, rather than attempt to fit both with a single evolution, and secondly,  by decoupling the amount of gas present in the system from the star formation rate. More explicitly, our previous model consisted of a gas reservoir that was used to form stars which then release gas and metals into the reservoir according to the chosen initial mass function (IMF) and selected stellar yields. Inflows and outflows were treated implicitly -- the gas reservoir was used only to dilute metals and form stars, there is no explicit information as to whether the gas was in the hot, warm or cold phase, and, indeed, could be in any form. We started the simulation with a total gas mass of unity, and constrained the system so that the integral of the star formation history is also unity. This meant that in normalised units, the stellar mass  was always less than unity, because of the gas returned to the ISM by each stellar population, mainly from AGB stars and SNII \citep[as discussed in ][]{Snaith2015}. 

We know, however, from detailed studies of galaxy formation and evolution that gas falls into galaxies over time \citep[e.g.][]{Larson1972, Dekel2006}. There is also a strong relationship between the cold gas density and the star formation rate \citep{Schmidt1959}. In most chemical evolution models it  is assumed that the amount of gas in the cold phase follows a Schmidt-Kennicutt law \citep{Schmidt1959,Kennicutt1998},

\begin{equation}
\log(\Sigma_g) =\log(\Sigma_{SFR})/k  - \log(\epsilon)
\label{EqnSFR}
\end{equation}
\noindent
where $\Sigma_g$ is the surface density of the star forming gas, $\epsilon$ is the normalisation constant, $k$ is a constant, usually assumed to be around k=1.5 \citep{Kennicutt1998}, and $\Sigma_{SFR}$ is the star formation rate surface density. Many models neglect gas that might be present in the warm or hot gas phases, and assume that all gas is potentially star forming. 

Our model makes use of the instantaneous mixing approximation\footnote{not to be confused with the instantaneous recycling approximation, where metals are immediately returned to the ISM.}, which is where we assumed that the metals mix with all the gas present at the time of formation. For a detailed discussion of the model see \citet{Snaith2015}.\footnote{in \cite{Snaith2015} we incorrectly quote the parameters for the fit of stellar lifetimes by \citet{Raiteri1996} for $a_1$ and $a_2$ but they are used correctly in the code itself. } 

In this paper we elaborate on our previous model to include explicit infall and a Schmidt-Kennicutt relation between gas surface density, $\Sigma_g$, and star formation rate surface density, $\Sigma_{SFR}$. The initial mass of reservoir is zero, and pristine gas is added over time according to a given infall law. We assume pristine gas, following previous work, such as \citet{Troncoso2014}. The exact mix of the infalling gas is unknown, and could be expected to be evolving. Thus, assuming pristine gas is a reasonable compromise, to reduce the number of parameters. Various authors have discussed pre-enrichment of infalling gas, such as \citet{Tsujimoto2010}. However, our assumption of pristine infall is a reasonable one, which is common in the literature considering the other uncertainties in the model \citep[e.g. stellar yields or IMF, ][]{Romano2010, Snaith2015}. 

We use two different formalisms for the infall law. In one set of runs the infall rate of pristine gas follows the traditional form, e.g. \citet{Chiappini1997}, the gas infall rate declines exponentially, and is determined by,
\begin{equation}
\Psi_{bary} = \sum^{n}_{i=1} \psi_i,
\label{Eqn:inf}
\end{equation}
\noindent
where,

\begin{equation}
\psi_i = \begin{cases}
0 &t<t_{start,i}\\
A_i \exp \left(\frac{-(t-t_{start,i})}{\tau_i}\right) & t \ge t_{start,i}\\
\end{cases}
\label{Eqn:inf}
\end{equation}
\noindent
where $n$ is the number of infall episodes, $t$ is the time, $t_{start,i}$ is the start of the infall episode, $\tau_i$ is the time scale of the episode. Each era of the model run is able to have its own infall rate and constant, $\epsilon_i$. For a given infall event, the total surface density after 14 Gyr is given by,
\begin{equation}
\Sigma_{bary}(t=14) = A\tau \left(1-\exp \left( \frac{-(14-t_{start})}{\tau} \right) \right),
\end{equation}
\noindent
In each time step we add pristine gas to the model with mass $\Psi_{bary} dt$.  We set the model so that the final baryonic mass surface density of the disc is 60 $M_\odot/pc^2$. 

Alternatively, we add gas without an explicit functional form. We subdivide the evolution into fourteen 1 Gyr eras with an infall rate given by the amplitude `A'. We make use of a modification of Eqn. \ref{Eqn:inf} and set $\tau$=1000 value and stop infall after 1 Gyr, e.g.

 \begin{equation}
\psi_i = \begin{cases}
0 &t<t_{start,i}\\
A_i \exp \left(\frac{-(t-t_{start,i})}{\tau_i}\right) & t_{start,i} < t \le t_{start,i}+1\\
0 &  t \ge t_{start,i}+1\\
\end{cases}
\label{Eqn:inftophat}
\end{equation}
\noindent
This produces a series of essentially `top hat' infall episodes, which sum together to form the overall infall history of the model galaxy. 

We calculate the star formation in a given time step according to Eqn. \ref{EqnSFR}, and the gas is removed from the ISM. We do not incorporate the frequently implemented star formation threshold of 7 $M_\odot kpc^{-2}$ \citep[e.g.][]{Chiappini1997}. We use $k=1.5$, and $\epsilon$ is normalized by $\sqrt{\Sigma_{gas}}$, or more formally, scaled by $\Sigma_{gas}^{k-1}$  \citep[e.g.][]{Chiappini1997}. This scales the star formation rate surface density with total baryon surface density, and means the star formation law is effectively scale-free. This means that our chosen total surface density (60 $M_\odot/pc^2$) is irrelevant for choosing a chemical evolution and could be any value.  

We then cycle back over all preceding time steps to calculate the mass and abundances released from already formed stars. Over the 14 Gyrs of the run, stars return approximately 30\% of their mass back into the ISM, most within the first Gyr. In the infall model a fraction of the gas is transformed into stars at each step according to Eqn. \ref{EqnSFR}. This is controlled by $\epsilon$. In our exponential infall runs, each era can have a different $\epsilon$ value. While, in order to be able to use our fitting procedure, we fix the $\epsilon$ value at all times for the top hat runs. 

In this paper we make use of the same yield tables used in the fiducial model in \citet{Snaith2014,Snaith2015}. We assume that the stellar yields overwrite the elements already present in the star when the gas is released back into the ISM. In particular we use \citet{Nomoto2006} for SNII, \citet{Karakas2010} for AGB stars and \citet{Iwamoto1999} for SNIa and an IMF from \citet{Kroupa2001}. We also set the initial values of the primordial gas to be 0.75 hydrogen and 0.25 helium. When calculating the [X/H] value for a given element we use the average atomic mass in the Universe as an approximation to the specific mix of isotopes released from stars.  Although, not the most recent stellar yields in the Literature we make use of these in order to compare our result to our previous work \citep{Snaith2014, Snaith2015} and leave an exploration of the impact of additional yield tables to the future. 

In order to mimic the observed MDF and $\alpha$DF from APOGEE we produce a histogram of the [Fe/H] and [Si/Fe] track weighted by the stellar mass distribution. However, for each time step we add a random error, drawn from a Gaussian distribution with $\sigma$=0.02. We generate 30 histograms, with different random errors, and normalize the resulting histogram so the sum of all bins is equal to unity. We use a fixed random number seed for each run, so our results are reproducible on different runs of the model (except when bootstrapping to calculate uncertainties).

The MDF and $\alpha$DF are calculated by binning the points along the stellar evolution track (number of time steps) by [Fe/H] or [$\alpha$/Fe] and is given by,

\begin{equation}
M(/\alpha) DF_j = \frac{\sum^{n_j}_{i=1}M_j(i)}{\sum(M_{i,j})},
\label{Eqn:hist}
\end{equation}
where SFH$_j$ is the stellar mass in bin $j$, M$_j$(i) is the sum of all stars in the jth bin and M$_{i,j}$ is the sum over all stars in all bins. 

As well as fitting the time-[Si/Fe] distribution, as we did in \citet{Snaith2014, Snaith2015} we also fit the MDF and/or $\alpha$DF according to,

\begin{equation}
\label{Eqn:diffsqr}
        \begin{aligned}
F_{ij} = & \sum_{i=1}^{n_{star}}(O_{[Si/Fe]}-M_{[Si/Fe]})_i^2 +   \\
            & \omega_{MDF} \sum_{j=1}^{n_{bins}}(O_{MDF}-M_{MDF})_j^2 + \\
            & \omega_{\alpha DF} \sum_{j=1}^{n_{bins}}(O_{\alpha DF}-M_{\alpha DF})_j^2         
        \end{aligned}
\end{equation}        
\noindent
where $O_{[Si/Fe]}$ is the value of the observed [Si/Fe] at a given age and $M_{[Si/Fe]}$ is the model value, and $O_{MDF, \alpha DF}-M_{MDF, \alpha DF}$ is the difference between the normalised observed and model  MDFs ($\alpha$DFs) in each bin, and $\omega$ is a tuning parameter which weights the difference of the distribution functions relative to the difference of abundances. Our choice for $\omega$ will be discussed in the following section.\footnote{We ignore the error on the observations for the purposes of the fit.}  We also have the option to vary the $\epsilon$ parameter in Eqn. \ref{EqnSFR}. 

In order to define a model with varying $\epsilon$ we make use of the results of \citet{Tacconi2018} to explore the evolution of the star formation efficiency as the galaxy mass increases. In \citet{Tacconi2018} their Eqn 5. provides an evolution of the depletion time of the gas as a function of redshift and galaxy stellar mass and is quoted here,

\begin{equation}
\label{Eqn:diffsqr}
        \begin{aligned}
\log(t_{depl}) =  & A_t + B_t \log(1+z)  + C_t \log(\delta  MS) +  \\ 
                          & D_t \log(\delta M_*) + E_t \log(\delta R)
        \end{aligned}
\end{equation}        
\noindent
where $A_t$, $B_t$, $C_t$, $D_t$ and $E_t$ are constants, $z$ is the redshift, $M_*$ is the stellar mass, $\delta MS$ is given by $\delta MS = sSFR/sSFR_{MS}$, where $sSFR_{MS}$ is given by
 
\begin{equation}
        \begin{aligned}
\log(sSFR_{MS}) =&  (-0.16 - 0.026 t_c)  \times(\log(M_*)+0.025) - \\
&  (6.51 - 0.11t_c) + 9
        \end{aligned}
        \label{Eqn:S14}
\end{equation}        
\noindent
 where $t_c$ is given by,

\begin{equation}
        \begin{aligned}
\log(t_c) = & 1.143 -   1.026  \log(1+z)-  0.599 \log^2(1+z) +\\ 
                 & 0.528\log^3(1+z)
        \end{aligned}
        \label{Eqn:S14}
\end{equation}        
\noindent
$\delta M_*$=$M_*/5\times10^{10}$ and $\delta R$ is a function of the effective radius, which we shall ignore. From   \citet{Tacconi2018} Table 3 we use the best-fit parameters for the model with $\log(sSFR_{MS})$ taken from Eqn. \ref{Eqn:S14}, originally defined by \citet{Speagle2014}. We define a characteristic $\epsilon$ evolution using the star formation rate and stellar mass growth from the best fit to the $\omega_{MDF, \alpha DF}=900$ run for the flat infall initial conditions, which we will discuss in the following section. We fit the resulting evolving $\epsilon$ with a 3-order polynomial of the form 

\begin{equation}
k (t) = at^3 + bt^2 + ct + d
\label{Eqn:cubic}
\end{equation}
\noindent
where ($a$, $b$, $c$, $d$) = (-0.0013, 0.0379, -0.3102, -0.0378), and $k$ is the modifier on $\epsilon$ with time. In order to calculate the total stellar mass and star formation rate from the surface densities we assume that the disc has an exponential profile with a radial scale length of 2.5 kpc \citep{Porcel1998}, and that the Sun is 8 kpc from the galactic centre. We select the fixed value of 2.5 kpc to be illustrative of the range of possible values of the radial scale length. It is plausible that the different discs of the Milky Way have different scale lengths \citep[e.g.][]{Bovy2014}, however, we use a varying $\epsilon$ to be illustrative of a range of possible models rather than to be a precise fit due to the large number of additional parameters that would need to be taken into account. The impact of changing $\epsilon$ is similar to that of changing r$_d$, changing one parameter is sufficient to explore a range of models.

\begin{figure} 
	\centering
	\includegraphics[scale=.7, trim={3.3cm 12.5cm 5cm 1.5cm} , clip]{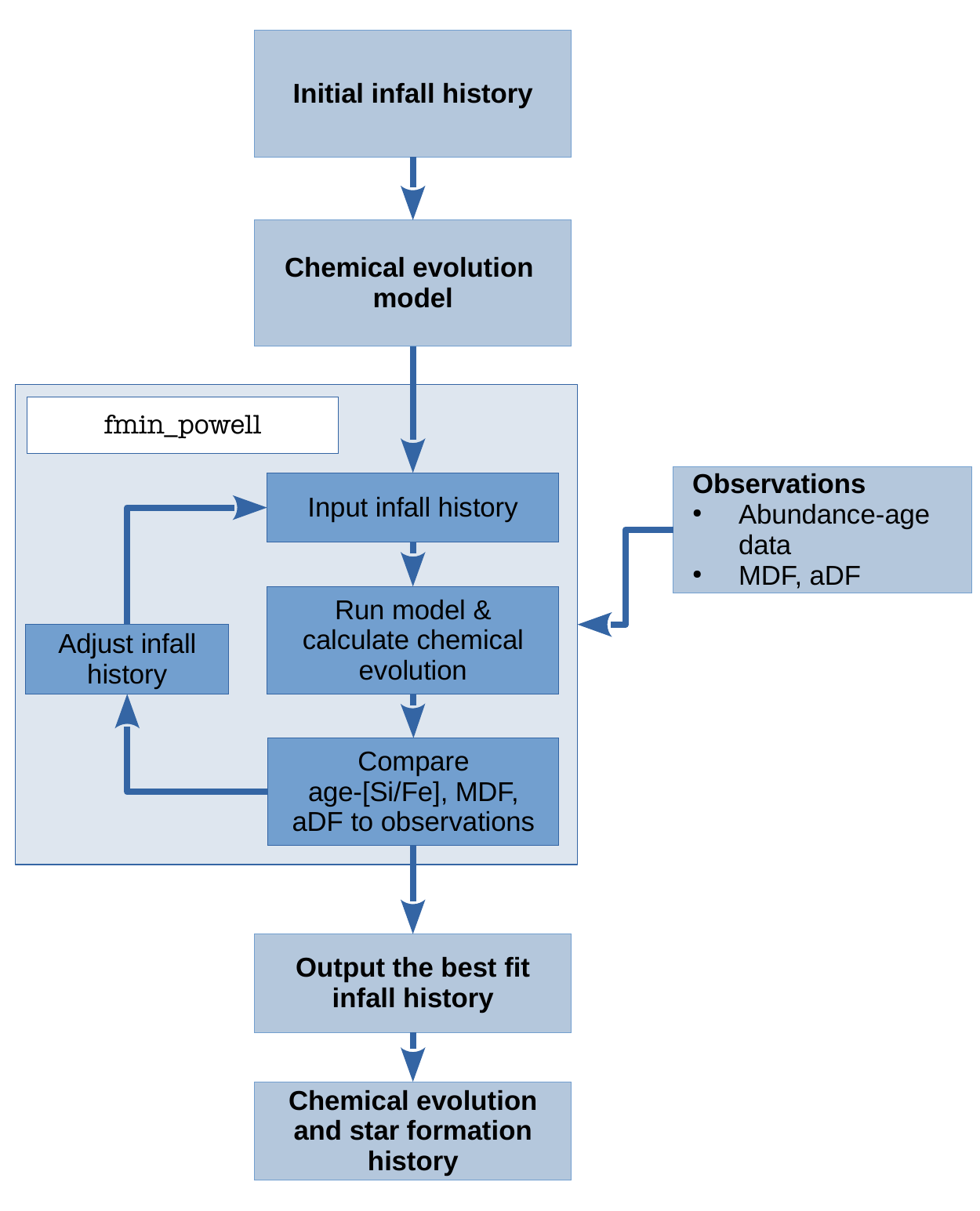}
	\caption{Schematic of the galactic chemical evolution model, and the algorithm to identify the best fit infall history.  }
	\label{fig:flowchart}
\end{figure}

Figure \ref{fig:flowchart} shows the fitting algorithm and how it interacts with the galactic chemical evolution model. We start with an initial infall history, which produces a chemical evolution track over the 14 Gyr of the evolution of the galaxy. We can compare this to the observations, using Eqn. \ref{Eqn:diffsqr}, to produce a single value out of the chemical evolution. We do this by comparing each star in the observations to the chemical evolution track according to their age. We also produce the MDF and $\alpha$DF using the same bins as the MDF and $\alpha$DF in the observations, following Eqn. \ref{Eqn:hist}. We make use of the fitting algorithm to minimize F$_{ij}$ by changing the parameters which define the infall history and $\epsilon$. We use the \texttt{fmin\_powell} method from  \texttt{scipy}  \citep{Virtanen2020} in  \texttt{Python}. 

\section{An infall history for the Milky Way}
\label{Sec:MainResults}

Exponentially decaying infall is the most usual form of infall given in the literature \citep[e.g. ][]{Chiappini1997,Spitoni2019}, so we first explore what exponential infall (as given by Eqn. \ref{Eqn:inf})  can tell us about the evolution of the Galaxy when fitted to the inner disc. We then explore the impact of adopting a more general formulation  for infall (given by Eqn. \ref{Eqn:inftophat})  on the build up of the disc.

\subsection{Model with exponential Infall}
\label{sec:expinf}
\begin{table*}
\begin{center}
  \begin{tabular}{ |c|cc|cccc|cc|cccc| } 
    \hline
    run & $A_1^+$ & $A_2$ &  t$_{s, 1}$ &  t$_{s, 2}$ &  t$_{s, 3}$ &  t$_{s, 4}$ &  $\tau_1^+$  &  $\tau_2^+$ & $\epsilon_1$ & $\epsilon_2$ & $\epsilon_3$ & $\epsilon_4$  \\
    \hline
    \hline
    long & 10 & 8 & 0 & 0.7 & 6 & 6.5 & 0.3 & 10 & 10 & 1.5 & 0. & 0.5\\
    medium & 20 & 12 & 0 & 0.7 & 6.0 & 7.23 & 0.1 & 5.0 & 8 & 1. & 0 & 0.4 \\
    short & 108 & 30 & 0 & 1 & 5 & 6.93 & 0.05 & 2 & 4.0 & 0.808 & 0.04 & 0.242  \\
    very short & 60 &30 & 0 & 1 & 5 & 6.93 &0.3 & 1.0 &  1.0 & 0.5 & 0.04 & 0.5 \\
    \hline
  \end{tabular}
\end{center}
\label{Tab:expoICs}
\caption{Initial parameters for the exponential runs. The values of `A' are renormalised so the total surface density of baryons is 60 $M_\odot/pc^2$.  The numbers on the parameters are: 1 -- onset, 2 -- thick disc, 3 -- quenching, 4 -- thin disc. Parameters which we fit are marked with a $+$. The various $\epsilon$s are allowed to scale according a constant.}
\end{table*}

We first divide the history of the galaxy into four epochs, corresponding to the onset, the thick disc, the `quenching era' and the thin disc era. The first era, the onset,  is the initial starburst and has a very rapid infall with a low $\tau$ and high `A'. This era acts to bring the [Si/Fe] down to the beginning of the data within the first Gyr. This is largely unconstrained, as we do not have data with ages greater than 13 Gyr while the model begins 14 Gyr ago. The second era is the second infall, which then builds up the stellar discs. The third era is the 'quenching' era we identified in \citet{Haywood2016}, and the fourth era is corresponds to the thin disc evolution. In correspondence to the \citet{Chiappini1997} GCE model the first and second eras are the first and second infall in that model, and the thirds and fours eras have no additional infall. 

Each era has a full set of parameters, including $A$, the amplitude of the infall, $t_{start}$, the onset of the infall, $\tau$ the timescale of the infall, and $\epsilon$, the star formation efficiency.  However, $A_{onset}$, $\tau_{onset}$ and $\epsilon_{onset}$ are fitted. $A_{thick}$ is set such that the final surface density of bayonic matter at z=0 is 60 M$_\odot/pc^2$, and $\tau_{thick}$ is varied from 2 to 10 Gyr (before fitting, and is one of the parameters fitted). There is, however, still infall occurring during the quenching and thin disc eras, due to the exponential tail of the infall. The star formation efficiency in each era is set to a constant value relative to one another, although they can vary globally.  The amplitude of infall during the quenching and thin disc phases are set to a very low and low value respectively.  Possible explanations for the quenching will be discussed in Section \ref{Sec:Discuss}. The quenching epoch was included to reproduce the strong change of gradient at around 8 Gyr in the [Si/Fe]-time distribution based on our experiences from \citet{Snaith2014,Snaith2015} and \citet{Haywood2016}. In those works, without the hiatus, the transition is more gradual, and would not replicate this key feature of the chemical evolution. The initial conditions are fitted by eye, then the overall model is fitted using \texttt{f\_min\_powell}. Although we included full quenching following previous work we will examine it in more detail in Appendix A.  They were fitted to roughly match the time-[Si/Fe] distribution first, and then to remove the early low [Fe/H] tail in the MDF. Removing this tail is mostly controlled by the parameters of the first infall era.  The key parameters for the four eras are given in Table 1. 

Where $\omega>0$ we fit the model using the time-abundance distribution, MDF and $\alpha$DF.  Where $\omega_{MDF}$=900 and $\omega_{\alpha DF}$=900 the distribution functions are strongly dominant over the time-[Si/Fe] distribution when fitting the data, whereas for lower values of $\omega_{MDF}$ and $\omega_{\alpha DF}$ (=0, 10) the time-[Si/Fe] distribution  is dominant over the metallicity and alpha-distributions. The use of high values of  $\omega$, such as 900, is advantageous  because ages are hard to calculate, and tend to have large uncertainties and other systematics, \citep{Leung2019}, but  has the disadvantage that it requires surveys with well understood selection functions.

\begin{figure*} 
  \centering
 \includegraphics[scale=.66]{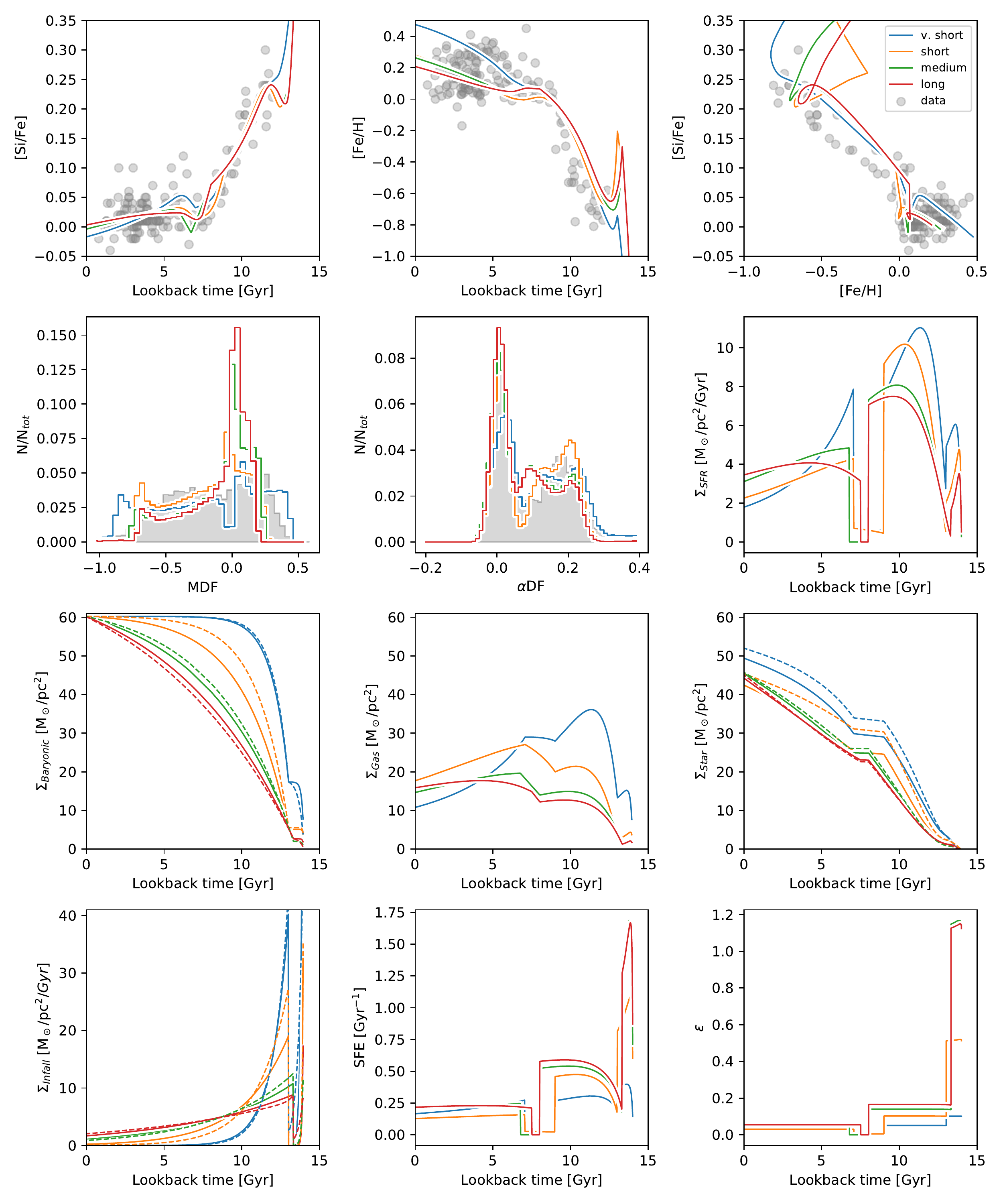}
  \caption{Fits of the exponential infall to the time-[Si/Fe] distribution ($\omega_{MDF,\alpha DF}$=0). The top row, from left to right show: the best-fit time-[Si/Fe] used to fit the data, the time-metallicity distribution, the abundance-metallicity distribution. In each panel the points show the data. The second row shows the MDF, $\alpha$DF and the recovered star formation history. The grey histograms show the distribution functions for the inner disc in APOGEE. The third row shows the evolution of the baryonic, gas and stellar mass over time. The fourth row shows the infall rate, star formation efficiency and $\epsilon$ evolutions. The different lines are for different initial conditions, and the dashed lines show the evolution of those unfitted initial states.}
  \label{fig:compareExpo}
\end{figure*}

\begin{figure*} 
  \centering
 \includegraphics[scale=.70]{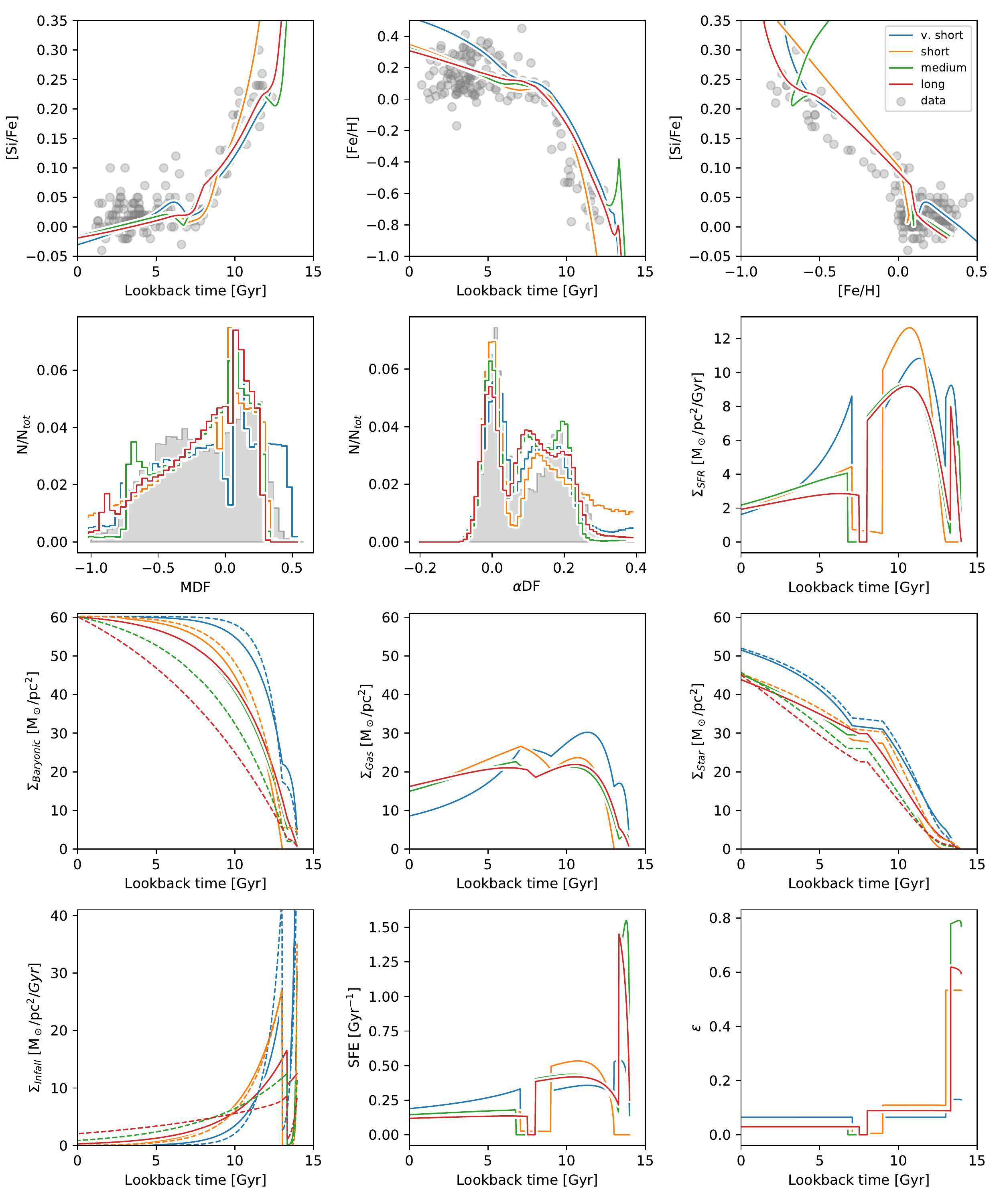}
  \caption{This shows the exponential infall model fitted to the MDF and $\alpha$DF as well as the time-[Si/Fe] with $\omega_{MDF}$=900 and $\omega_{ \alpha DF}$=900. The panels are the same as Fig. \ref{fig:compareExpo}.}
  \label{fig:compareExpo900}
\end{figure*}

Figures \ref{fig:compareExpo} and \ref{fig:compareExpo900}  show a number of fitted chemical evolutions using the exponential infall model. Figure  \ref{fig:compareExpo} shows the results fitted to the time-[Si/Fe] only ($\omega_{MDF}$ and $\omega_{\alpha DF}$ = 0.0).  The evolution of the initial guess is shown as dashed lines, while the fitted results are the solid lines. The fitting function does not significantly change the resulting infall history, and has a small effect on the overall star formation history. The different age-[Si/Fe]  evolutions can be fitted with a range of infall/$\epsilon$ values. However, as the initial infall time increases, the fit to the MDF becomes increasingly poor. 

\begin{table}
\begin{center}
  \begin{tabular}{ |c|ccccc| } 
    \hline
    run & $\alpha$DF & MDF & Abund & t$_{form, bary}$ &  t$_{form, stars}$  \\
   \hline
    \multicolumn{6}{|c|}{$\omega$=0.} \\
    \hline
long & 0.005 & 0.035 & 0.13 & 4.59 & 5.80 \\
medium & 0.003 & 0.016 & 0.13 & 3.96 & 5.57 \\
short & 0.004 & 0.005 & 0.11 & 2.72 & 4.49 \\
v. short & 0.006 & 0.007 & 0.14 & 1.36 & 4.24 \\
\hline
 \multicolumn{6}{|c|}{$\omega$=900.} \\
 \hline
long & 0.004 & 0.003 & 0.17 & 2.41 & 4.56 \\
medium & 0.003 & 0.003 & 0.15 & 2.62 & 4.76 \\
short & 0.004 & 0.003 & 0.41 & 2.52 & 4.32 \\
v. short & 0.003 & 0.006 & 0.18 & 1.33 & 4.10 \\
\hline
 \multicolumn{6}{|c|}{initial} \\
 \hline
long & 0.006 & 0.045 & 0.13 & 4.96 & 6.23 \\
medium & 0.004 & 0.012 & 0.15 & 3.65 & 5.41 \\
short & 0.006 & 0.003 & 0.18 & 2.17 & 3.96 \\
v. short & 0.004 & 0.010 & 0.19 & 1.33 & 3.86 \\
\hline
  \end{tabular}
\end{center}
\label{Tab:expomodelt}
\caption{The fitting function values for the $\alpha$DF, MDF and abundance-time distribution, along with the baryonic and stellar formation times, for the different exponential infall runs.}
\end{table}

However, when we set $\omega_{MDF, \alpha DF} = 900$ we see different behaviour. We see a much stronger effect on the fitted infall and star formation history, as shown in Fig. \ref{fig:compareExpo900}. The fitted $\tau$ values are pulled to much shorter timescales overall. This can be seen in Table 2. The fitted timescales are drawn to more rapid formation times (which is the time taken to build half the baryonic mass). If $\tau_{initial} > 2$ then $\tau_{fitted}$ is close to 2. Only the $\tau_{initial}$=1 has a shorter fitted timescale. The `short' run shows a greater deviation from the abundance-time distribution than the other runs, presumably due to the lack of an onset era in the recovered SFH. 

\begin{figure*} 
  \centering
  \includegraphics[scale=.55]{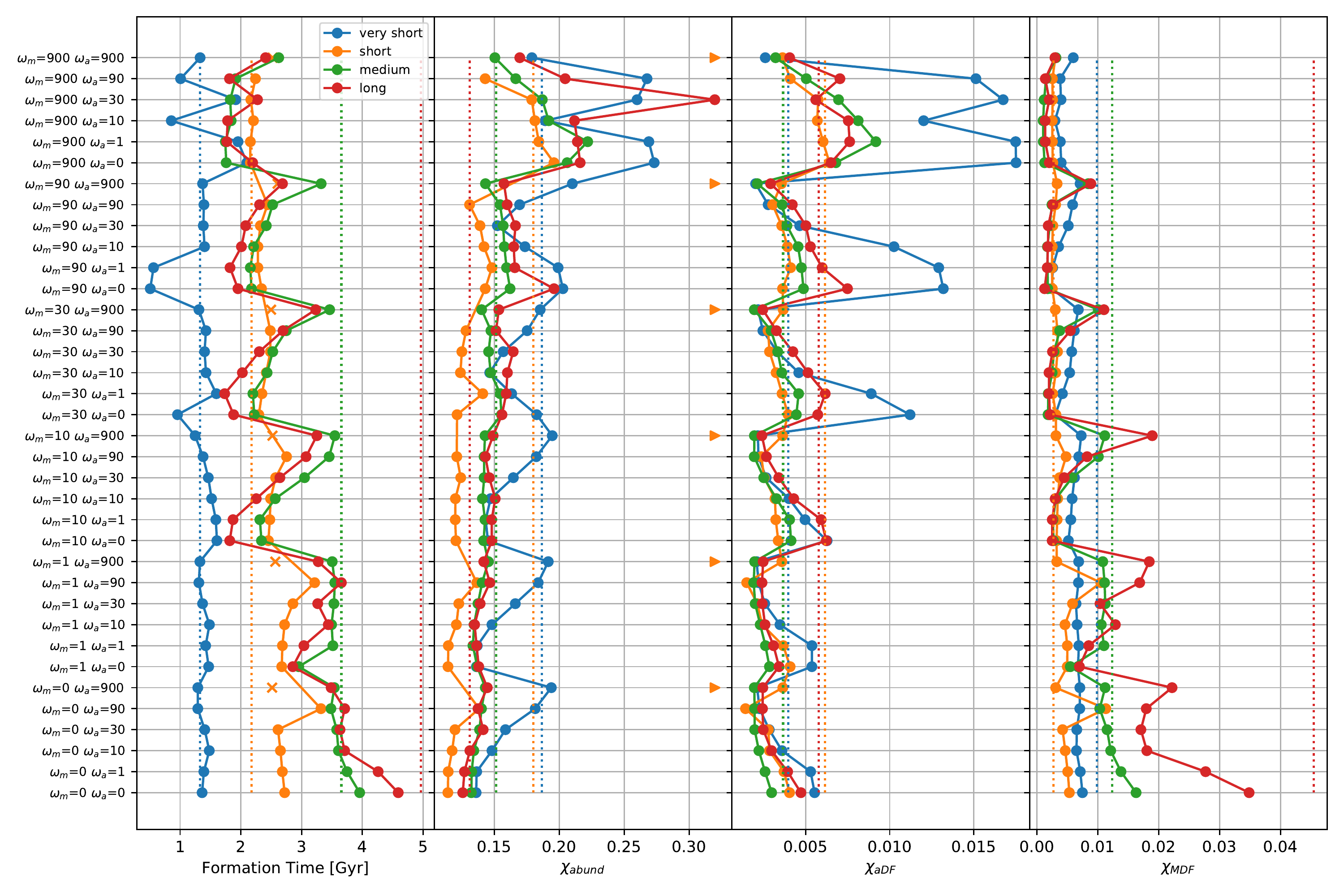}
  \caption{The formation time and fitting parameters for the exponential infall runs with different $\omega$ values. From left to right, the first  panel shows the formation time for the different fitted runs. The second panel shows the values for the fitting parameter to the abundance-age relation. The third panel shows the fitting parameter to the $\alpha$DF, and the fourth panel shows the fitting parameter to the MDF. The triangular points show failed runs, where the end result is larger than the ICs. In the left panel, a point is a triangle if any of the fitting parameters is larger than a threshold. The vertical dashed lines show the values for the initial conditions.}
  \label{fig:ExpFT}
\end{figure*}

The importance of the MDF in reproducing rapid infall no matter the initial parameters is shown in Fig. \ref{fig:ExpFT}. This figure shows that as the weighting given to the MDF decreases the infall timescale drifts to higher and higher formation times. The best-fit distribution is the one using  $\tau_{initial}$=2 Gyr for the high $\omega$ runs, with the MDF as the strongest constraint. Where the infall timescale is long, and the fit to the MDF and $\alpha$DF is low or zero, the high metallicity peak of the MDF builds up strongly until it is far in excess of the data. We see that adding a strong weighting to both DFs  fits the MDF, but this has a less strong effect on the $\alpha$DF. Indeed a stronger fit to the MDF comes at the expense of the fit to the $\alpha$DF. The dip in particular is not significantly better fit with $\omega_{MDF, \alpha DF}$=900 than with $\omega_{MDF, \alpha DF}$=0, except for the `long' ICs. Thus, when using high $\omega$ values, the improvement to the fit is strongest for the MDF, but at the price that the $\alpha$DF is not much improved.

Overall, this suggests that the data (when fitted to abundances, $\alpha$DF and MDF) favour rapid early accretion in the inner disc, regardless of the initial timescale before fitting. Similarly, we see that the fit to the abundance-time distribution alone is degenerate. The model struggles in one particular respect: the dual peak distribution of the MDF is poorly recovered in many instances of the exponential model, even with the quenching epoch. The $\alpha$DF also shows a third peak for the long infall run, suggesting that the fit is not perfect. This may be due to the small number of degrees of freedom, which prevents careful tailoring of the SFH to the data. This is because the parameters to be fitted were limited so that the parameter space was manageable.

 Different combinations of infall and star formation efficiency may exist that could fit the data better, which we did not find.  Our results are clearly indicative of a rapid infall, given that where the evolution is strongly constrained each run shows a very similar rapid evolution. The MDF is the stronger determinator, and harder for the model to closely fit. This could be due to the MDF being less responsive to the the evolution of the system than the $\alpha$DF. The $\alpha$DF is more responsive to the instantaneous SFR, due to the very short time scales of SNII. This was seen in, for example, \citet{Snaith2016}, where a strong local starburst can result in a strong peak in the [$\alpha$/FE] evolution. Meanwhile, the [Fe/H] evolution is a more complex function of the stellar yields for SNII or SNIa, the SNIa time delay function, the IMF etc. Thus, the tools we have available to control the evolution, the SFE and the infall rate, is not sufficient on their own to fully reproduce the MDF.  In addition, the  difference in the model and observed DFs could be influenced by the Milky Way bar. The relative absence of low-Fe , high-$\alpha$ stars in the model may be due to contamination by stars in the bar (which is 4.5--5 kpc long). As we have no dynamical information in the model, this is cannot be taken into account. Similarly, we might expect that the fraction of thin-disc, metal rich and kinematically cold stars would be higher at the edge of the bar.

\subsection{Non-parametric infall model}

Unlike in traditional infall models described by Eqn. \ref{Eqn:inf}, we opt for the more flexible `top hat' infall formulation given by Eqn. \ref{Eqn:inftophat}. We break the 14 Gyr of the model's evolution into fourteen 1 Gyr bins, with each era following Eqn. \ref{Eqn:inftophat}.  We fix $\epsilon$ for each infall episode and do not allow it to vary relative to each other, except where predetermined (for example using Eqn. \ref{Eqn:cubic}, or when quenched). The fitting function acts on the amplitude of the infall in each bin, and the global normalisation of $\epsilon$. Thus, the global infall is the sum of the fourteen independent infall eras. This allows us to be more flexible about the infall history than when we assumed the exponential functional form.

We run several scenarios to explore the influence of the specifics of the model, and the initial conditions, on the star formation history and chemical evolution. We fit the model using the time-abundance  distribution, and optionally the MDF and $\alpha$DF.

The five different types of run we use for the `top hat' infall model involve different initial conditions and variations to $\epsilon$. These include:
\begin{table}
\begin{tabular}{ llll } 
\hline
model & name              & input infall      & epsilon(t)         \\
\hline
1 & flat           & flat              & constant           \\
2 & high $\omega$         & output of model 1 & constant           \\
3 & flat, vary $\epsilon$   & flat              & variable           \\
4 & high $\omega$ , vary e & output of model 1 & variable           \\
5 & quench         & output of model 1 & quenched\\
\hline
\end{tabular}
\label{Tab:thesemodels}
\caption{Description for the models for the non-parametric runs}
\end{table}

\begin{enumerate}
\item flat -- the infall in the initial conditions is constant across cosmic time before the fitting function is applied,
\item high $\omega$ -- a run fitting the already fitted result of run 1 (with $\omega_{MDF, \alpha DF}$=900). This was done for later comparisons with other values of $\omega_{MDF, \alpha DF}$
\item flat, vary $\epsilon$ --  flat infall initial conditions but an $\epsilon$ which varies according to Eqn. \ref{Eqn:cubic},
\item high $\omega$ , vary $\epsilon$ --  a run fitting the already fitted result of run 1 but an $\epsilon$ which varies according to Eqn. \ref{Eqn:cubic},
\item quench -- quenched star formation with infall ICs from the best-fit run of run 1. We apply quenching, such that the $\epsilon$ is almost zero for 3 Gyr around 8 Gyr ago.
\end{enumerate}

\noindent
This is summarised in  Table 3. We allow $\epsilon$  to vary according to Eqn. \ref{Eqn:cubic} in order to explore the importance of this parameter on the chemical evolution. This expected to be most important at early times.

\subsubsection{Constraining with the local data only}
\label{Sec:OtherResults}

\begin{figure*} 
  \centering
 \includegraphics[scale=.68]{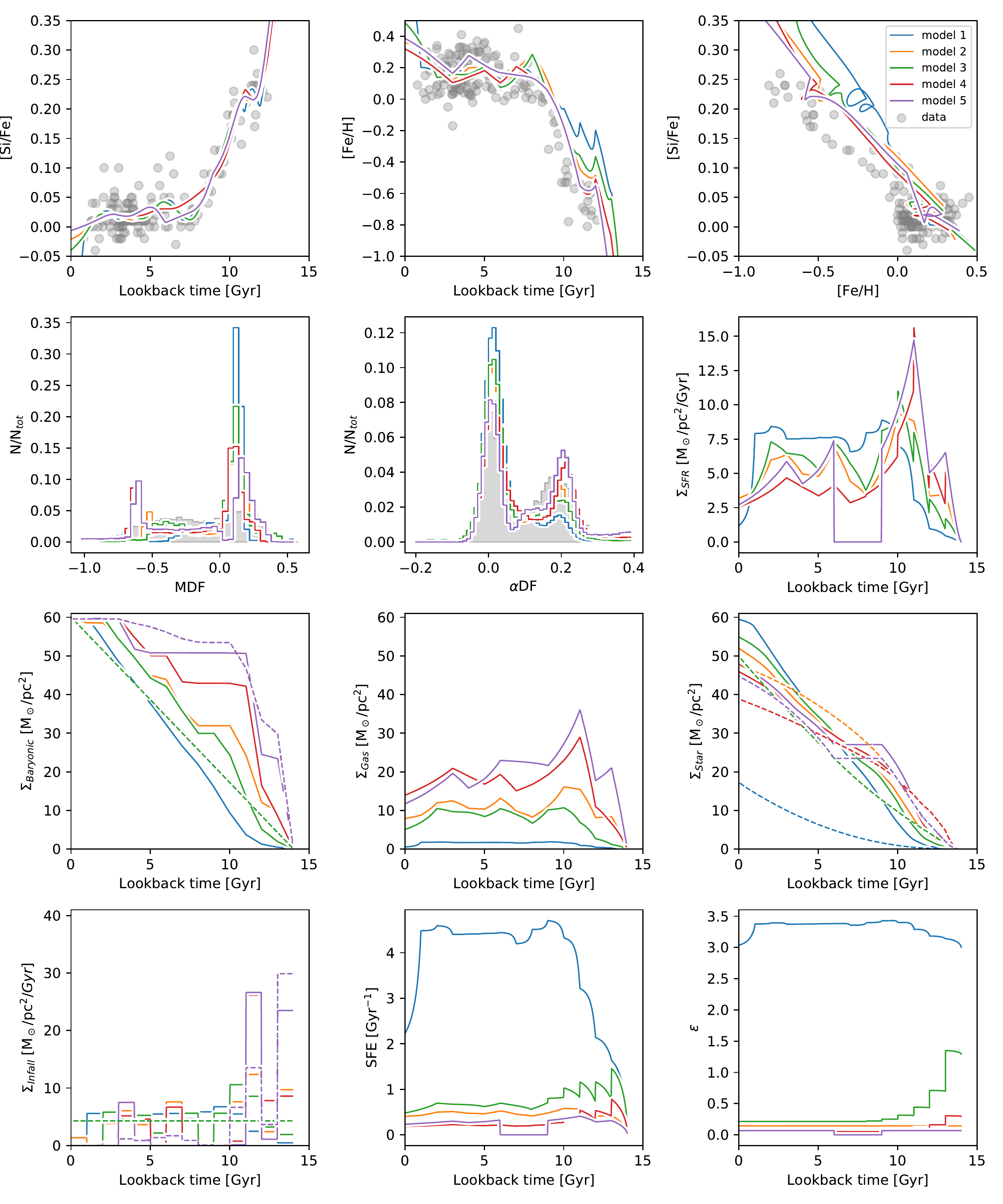}
  \caption{This shows the free infall model fitted to the time-[Si/Fe] with $\omega_{MDF}$=0 and $\omega_{\alpha DF}$=0. The panels are the same as Fig. \ref{fig:compareExpo}. The different lines show different runs, where the model numbers are referenced in Table 3. }
  \label{fig:Free0}
\end{figure*}

In this section we discuss the free infall model fitted only to the abundance-time distribution (e.g. where $\omega_{MDF}$=0. and $\omega_{\alpha DF}$=0.). This will allow us to assess the importance of including the distribution functions in our fit. Formerly, we fitted our closed box model to only the time-[Si/Fe] distribution, and found that the resulting star formation history and chemical evolution fitted the MDF and $\alpha$DF well. This was effectively a prediction of the model and was interpreted as a success for our derived star formation history \citep{Snaith2015}. 

 Figure \ref{fig:Free0} shows that when fitting just the time-[Si/Fe] distribution we are able to achieve an excellent fit without the need for explicit quenching. We also fit the [Fe/H]-time distribution adequately, although the flat initial condition run increases in [Fe/H] too rapidly at early times. However these fits come at the expense of the fit to the MDF (which  is an order of magnitude worse than for the high $\omega$ runs). Although we have a reasonable fit to the $\alpha$DF, the fit to the MDF peaks very strongly at high metallicity. This implies that the model predicts too much star formation at later times. This can be seen in the star formation history panel, where the star formation rate is higher at later times compared to the fits using the distribution functions.  Table 4 shows that the median baryonic accretion timescales are longer than in the exponential models (with high $\omega$ ) where $\omega$=0, ranging from  6.5 Gyr to 2.3 Gyr. As previously mentioned, the different responsiveness of the two DFs to the infall is due to the different origins of the $\alpha$ and Fe elements. This is one of the strengths of using the two types of elements in the first place, because their different origins and time scales provides access to different processes. The MDF responds less well to the current $\Phi_{gas}$ or SFE than to other details such as the IMF, yields, SNIa time delay function, binary fraction etc.

 The infall histories unconstrained by the distribution functions show secondary infall at later times assembling over 2-7 Gyr. The quenching era fully constrains the infall, bringing the assembly time in line with the high sigma runs.  This suggests that the inner disc is in place early and did not experience much accretion at later times even though star formation was ongoing.

\begin{table*}
\begin{center}
	\begin{tabular}{ cllllll } 
		\hline

scenario & run & abundance	&	$\alpha$DF	&	MDF	&	T$_{bary}$	&	  T$_{star}$	\\
\hline
		 \multicolumn{6}{c}{$\omega_{MDF, \alpha DF} = 900$} \\
\hline
1 &flat & 0.18	$\pm$ 0.03 &	0.0065	$\pm$ 0.0006 &	0.0027	$\pm$ 0.0004 &	0.77	$\pm$ 0.1 &4.8	$\pm$ 0.1 \\
2 & high $\omega$& 0.19	$\pm$ 0.02 &	0.0047	$\pm$ 0.0003 &	0.0030	$\pm$ 0.0002 &	2.00	$\pm$ 0.3 & 5.2	$\pm$ 0.1 \\
3 & flat, vary $\epsilon$ & 0.65	$\pm$ 0.5 &	0.0047	$\pm$ 0.003   &	0.0029	$\pm$ 0.002   &	3.70	$\pm$ 0.8 &6.2	$\pm$ 0.6 \\
4 & high $\omega$, vary $\epsilon$ & 0.19	$\pm$ 0.04 &	0.0041	$\pm$ 0.0005 &	0.0023	$\pm$ 0.0005 &	1.90	$\pm$ 0.3 &4.3	$\pm$ 0.4 \\
5 & high $\omega$, quench & 0.17	$\pm$ 0.02 &	0.0016	$\pm$ 0.0002 &	0.0045	$\pm$ 0.0003 &	1.70	$\pm$ 0.2 & 4.3	$\pm$ 0.08 \\
\hline
 \multicolumn{6}{c}{$\omega_{MDF, \alpha DF} = 0$} \\
\hline
1 & flat & 0.11	$\pm$ 0.02 &	0.0120	$\pm$ 0.003 &	0.055	$\pm$ 0.02 &	6.5	$\pm$ 0.4 &	7.8	$\pm$ 0.2 \\
2 & high $\omega$&0.11	$\pm$ 0.01 &	0.0086	$\pm$ 0.002 &	0.032	$\pm$ 0.01 &	3.4	$\pm$ 0.8 &	7.0	$\pm$ 0.4 \\
3 & flat, vary $\epsilon$ & 0.11	$\pm$ 0.02 &	0.0100	$\pm$ 0.002 &	0.039	$\pm$ 0.02 &	5.9	$\pm$ 1 &	7.5	$\pm$ 0.2 \\
4 & high $\omega$, vary $\epsilon$ &0.12	$\pm$ 0.02 &	0.0061	$\pm$ 0.002 &	0.024	$\pm$ 0.007 &	2.6	$\pm$ 0.4 &	5.4	$\pm$ 0.6 \\
5 & high $\omega$, quench &0.12	$\pm$ 0.01 &	0.0048	$\pm$ 0.002 &	0.024	$\pm$ 0.007 &	2.3	$\pm$ 0.3 &	4.6	$\pm$ 0.2 \\
\hline
\end{tabular}
\label{Tab:allruns2}
\caption{Fitting parameter values for the different distributions for the free infall runs shown in Figs. \ref{fig:Free0} and \ref{fig:Free900}. The columns `abundance', '$\alpha$DF' and `MDF' show the different fitting parameters, and T$_{bary}$ is the median formation time for the baryons from 24 bootstrapped samples, whilst  T$_{star}$ is the formation time for the stars. The errors are the standard deviation of the 24 bootstrap runs. }
\end{center}
\end{table*}

\subsubsection{Adding the inner disc MDF and $\alpha$DF}

\begin{figure*} 
\centering
\includegraphics[scale=.65]{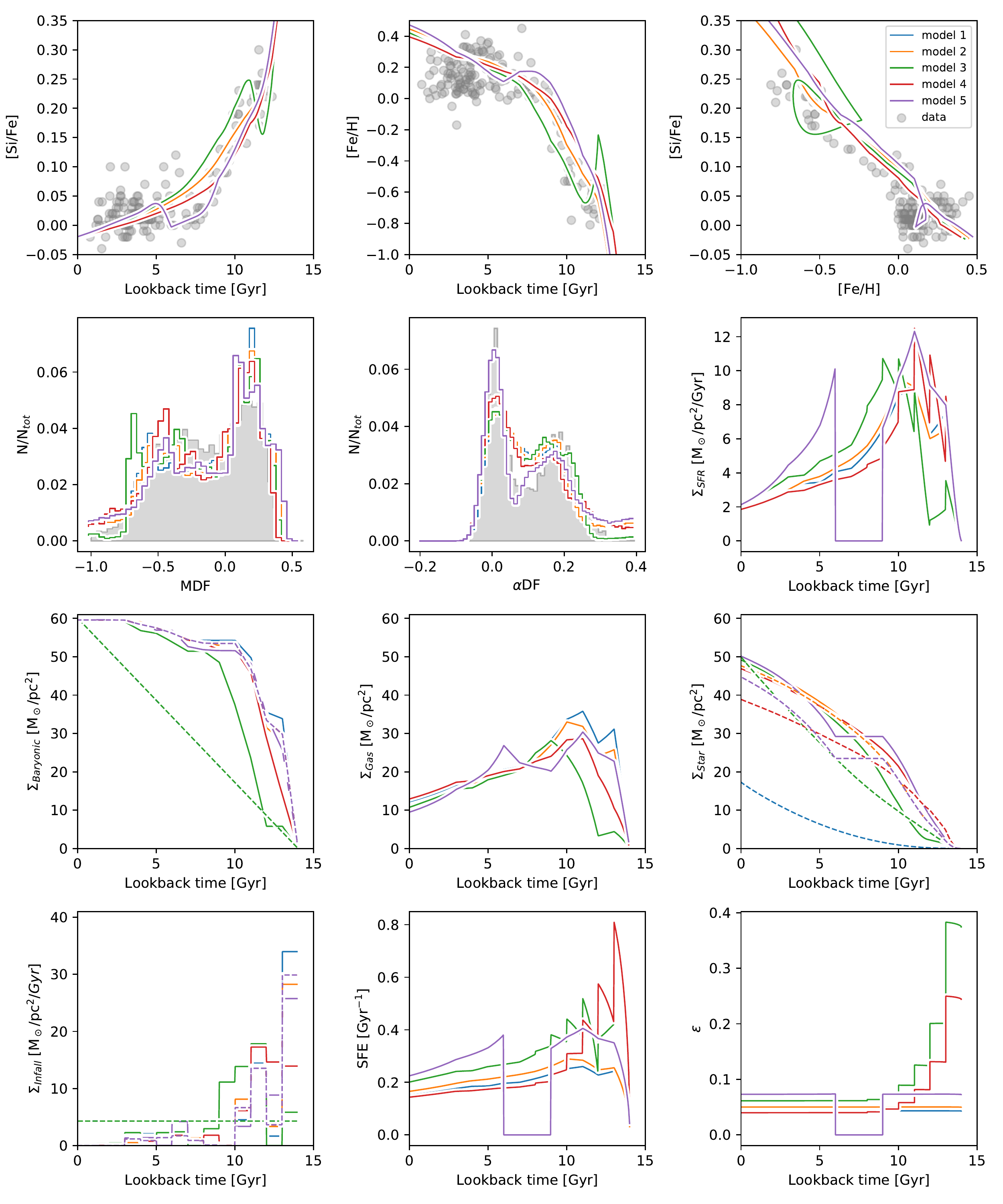}
\caption{This shows the free infall model fitted to the MDF and $\alpha$DF as well as the time-[Si/Fe] with $\omega_{MDF}$=900 and $\omega_{\alpha DF}$=900. The panels are the same as in Figure \ref{fig:compareExpo}.}
\label{fig:Free900}
\end{figure*}

\begin{figure*} 
\centering
\includegraphics[scale=.60]{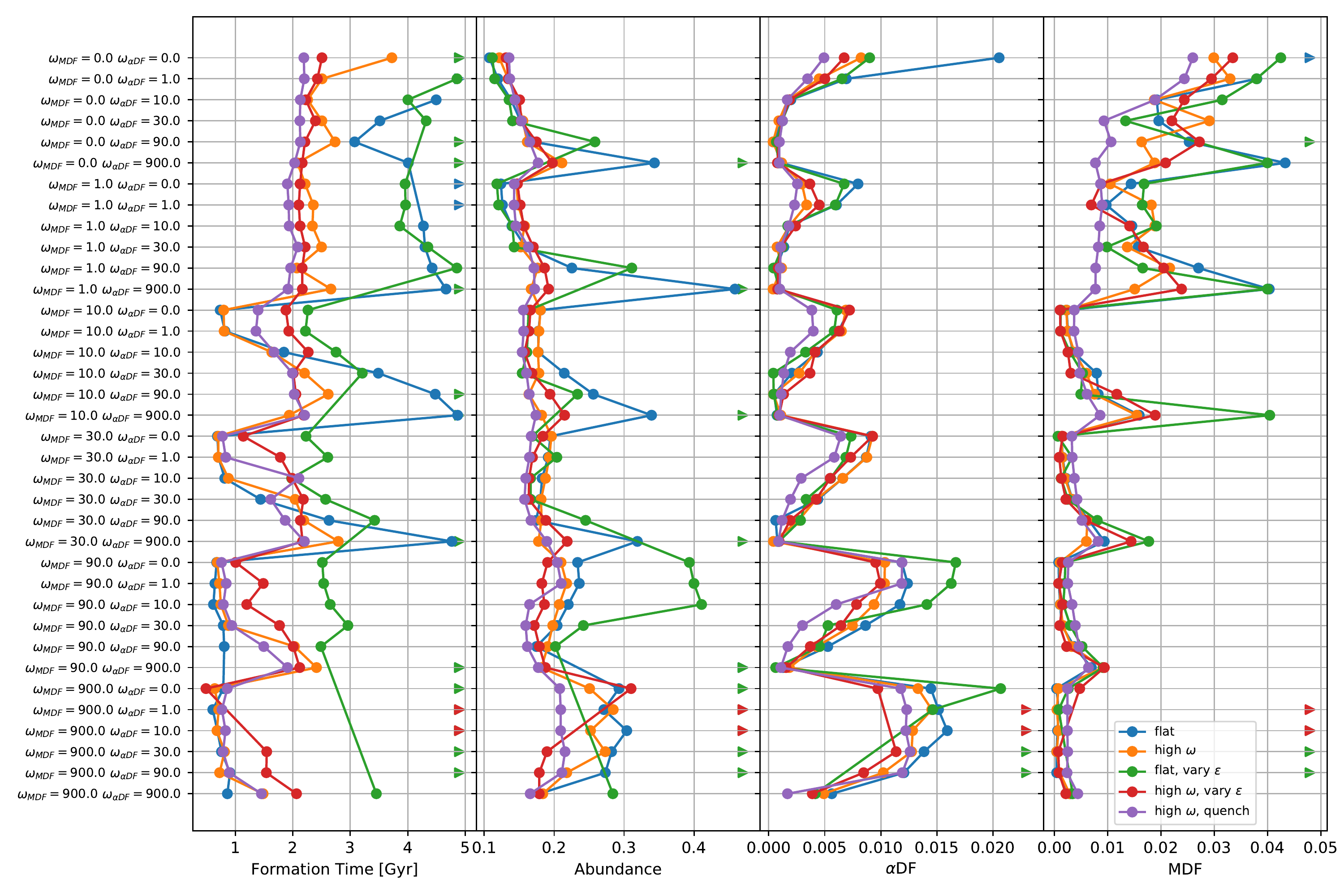}
\caption{The formation time and fitting parameters for the free infall runs with different $\omega$ values. From left to right, the first  panel shows the formation time for the different fitted runs. The second panel shows the values for the fitting parameter to the abundance-age relation. The third panel shows the fitting parameter to the $\alpha$DF, and the fourth panel shows the fitting parameter to the MDF. The triangular points show failed runs, where the end result is larger than the ICs. In the left panel, a point is a triangle if any of the fitting parameters is larger than a threshold. }
\label{fig:FreeFT}
\end{figure*}

Figure \ref{fig:Free900} shows the best-fit infall, star formation and chemical evolution for each of our five best-fit runs with $\omega_{\alpha DF, MDF}$=900. The blue and green lines show the results of runs from the flat infall initial conditions but the runs shown in orange, red and purple lines use the result shown by the blue line as its initial condition.

We used a higher initial $\epsilon$ normalisation value for the runs where we make use of Eqn. \ref{Eqn:cubic} (scenarios 3 and 4), in order to improve convergence, because without the renormalisation the function minimisation method struggles to fit the data. This change is to compensate for the decrement resulting from the application of Eqn. \ref{Eqn:cubic} to the $\epsilon$. The figure shows the best-fit chemical evolution history (top row), and the distribution functions (left and middle panels on the second row) for each of the four models, along with the corresponding star formation history (right hand panel on the second row). The baryon, gas and stellar surface density evolutions are shown on the third row, and the infall rate, star formation efficiency and $\epsilon$ are shown on the bottom row. 

Each of the models is a reasonably good fit to  the MDF, $\alpha$DF and time-[Si/Fe], as they should, considering they were fitted to these distributions. They also match well with the time-[Fe/H] and [Fe/H]-[Si/Fe] distributions. However, it is noticeable that the plots without either explicit quenching or a strong change in $\epsilon$ do not show a strong gradient change at around 8 Gyr, as can seen in the data, as well as in our work using the closed box model. We previously found  \citep[in][]{Snaith2014, Snaith2015, Haywood2019},  that a star formation hiatus is able to produce the gradient change. We therefore included scenario 5 to reproduce this. The sharp gradient change is clear in scenario 5, although it may be that the quenching epoch is too long for an ideal match to data. This long quenching era  would lead to a hiatus in the age-abundance distribution around 8 Gyr, which is not obvious from the observations. The other evolutions show a too gradual change of gradient, however, to reproduce this key feature. 
 
 This implies that there may be additional physics shaping the chemical evolution of the Milky Way beyond the connection between infall and star formation rate. This is because in our model there is plenty of gas remaining in the system when the quenching occurs, meaning that the quenching cannot be just the result of the ISM running out of material to fuel star formation \citep[as is used in other GCE models such as ][]{Spitoni2019}. We will discuss possible scenarios for this in Section \ref{Sec:Discuss}. Within the context of this model, the infall and star formation are not enough to fully reproduce the chemical evolution. As far as the model can be expected to reasonably reproduce the properties of the Milky Way, we would need to include  an alternative method of quenching.  

Table 4 presents the value of the fitting parameter to the MDF, $\alpha$DF and abundance-time distributions for our best-fit runs, along with the predicted formation time of the disc (which is where half the final mass of the disc is in place).  The uncertainties are calculated from best-fits to 24 bootstrapped samples of the observational datasets and different random errors used to generate the distribution functions. The values are the medians of the bootstrapped runs, and the errors are their standard deviations. The best fits to the abundance-time distribution is the quenching run, although the difference between the runs is small and well within the error associated with bootstrapping the data. The worse fit to the abundance-time distribution is the run with varying $\epsilon$ and flat initial infall, which is a significant outlier in terms of the baryon formation time. However, there is a corresponding increase in the model uncertainty, meaning that formally, there is no meaningful difference in the fits to the time-abundance distribution. The large error does indicate that this model is the least successful. This run does, however, show a good match to the $\alpha$DF and MDF. 

We also see that the recovered infall of the inner disc strongly favours the rapid early accretion of matter (within approximately the first 2 Gyr). In each run, the time taken for the galaxy to assemble half its baryonic mass is around 2 Gyr. This suggests that most of the inner disc is in place at early times. There is a pause in infall after around 10 Gyr in each run, except for scenario 3, and some subsequent infall, although only a small amount ($<$10  $M_\odot/pc^2$). Star formation occurs more slowly, with an assembly time of 4--5 Gyr (co-incident with the thick disc formation era). This also matches well with the closed box model from our previous work, where the thick disc is a massive component of the Galaxy. The predicted formation times are very similar within the error, except for the run with varying $\epsilon$ with a flat initial infall, which has a longer timescale. This is likely due to the fitting procedure as the other runs agree well, and the error on this run is much larger. 

We also note that the [Fe/H]-[Si/Fe] evolution of the different runs show a similar evolution. The chemical evolution track between [Fe/H] =-0.5 to 0 lies above the  \citet{DelgadoMena2017} data. However, the different evolutions very closely conform to the data selected from APOGEE. This should not come as a surprise as the models are most strongly constrained by the two distribution functions.

In order to understand the impact of $\omega$ on the resulting fits we explored a range of values. Figure \ref{fig:FreeFT} shows the impact of sigma on the formation time and fitting parameter to the different observables. We see that where the $\omega_{MDF}$ is high the formation time is very low, but where $\omega_{\alpha DF}$ is higher than  $\omega_{MDF}$ there is a slight push to longer formation times. The figure also shows that the fitting parameter for the abundance-time distribution is strongly dependent on the values of $\omega$, with the fitting parameter rapidly growing as the distribution functions make an increasing contribution to the overall fit. At higher weighting, each run, excluding scenario 3, are very closely matched with rapid formation times. For the run with an initially flat infall the star formation efficiency is very high, and the subsequent gas surface density at the present time is very low, suggesting that the flat ICs are too divergent from the final result to avoid unphysical local minima in parameter space. The other distributions have more reasonable surface densities for the gas and stars. Overall, when quenching is explicitly included, the infall is quite constrained, and tends towards shorter infall timescales and realistic gas fractions.  As before, the fit to the dip in the $\alpha$DF is not improved, and even made worse, in the high $\omega_{MDF}$ runs. This is the result of the improved fit to the MDF. This suggests that improvements should be made to the yields, IMF, SNIa delay time etc., to better fit both DFs at the same time. Even if the fit is relatively imprecise, the overall behaviour is clearly captured, suggesting that even if our fit is not perfect, it is at least identifying all the key features.

Each run with high $\omega$, and many of the runs with $\omega$=0 (the flat infall run is the notable exception because of a very high value of $\epsilon$) produce a similar accretion history and a local stellar surface density of around 50 $M_\odot/pc^2$, and a gas density of 10-15 $M_\odot/pc^2$, which is within the range of observations \citep{Bovy2013, Zhang2013}, if a little gas-poor for the solar vicinity.

\begin{figure} 
	\centering
	\includegraphics[scale=.55]{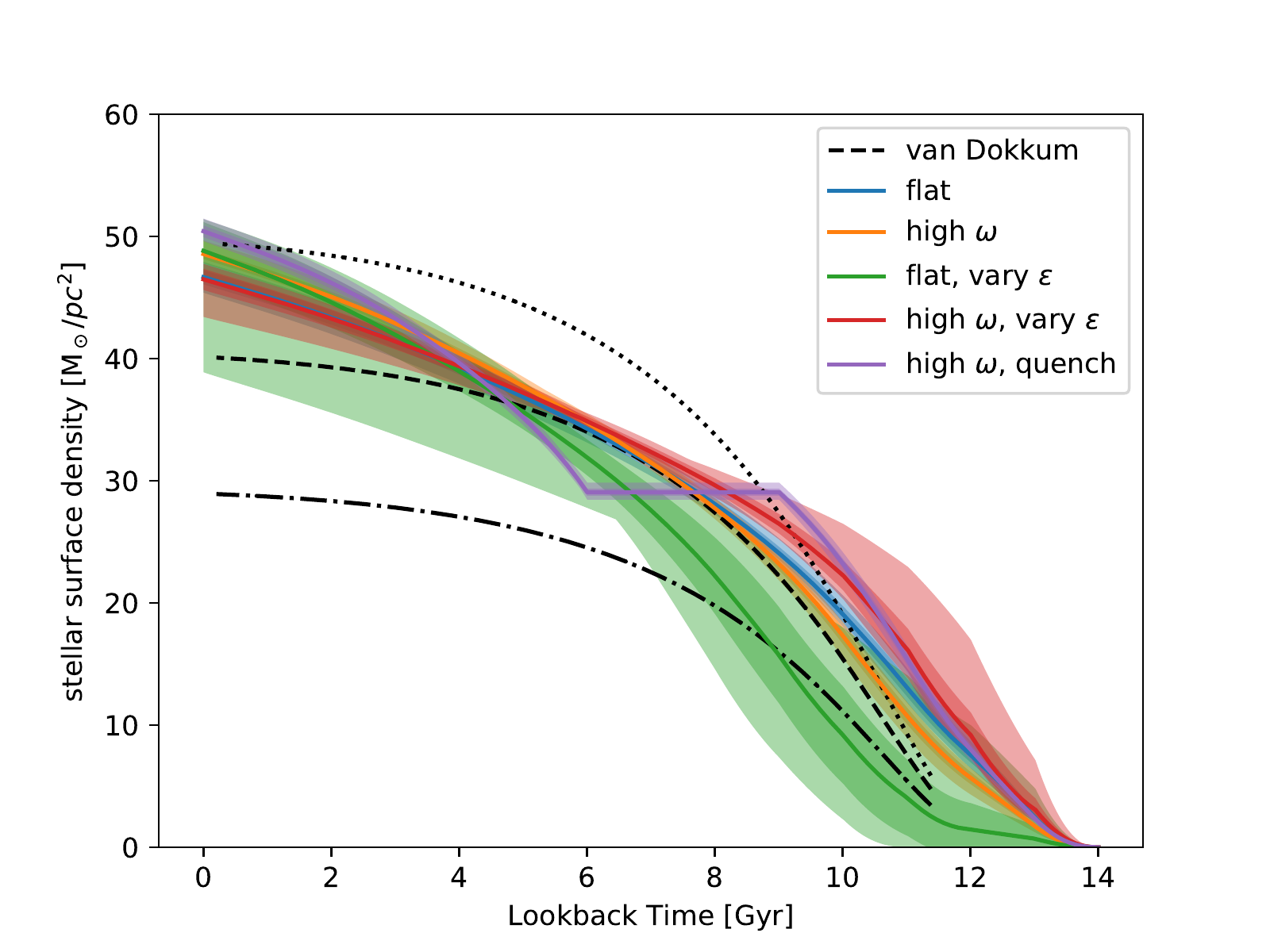}
	\caption{Comparison between the models with free infall and high $\omega$ compared to the average stellar mass growth and star formation history from \citet{vanDokkum2013}. We also show the median line of 24 runs where we bootstrapped the data. The variation from the minimum value to maximum value in each time-step is shown as the light coloured region around each line. $\pm$ the standard deviation is shown as the darker coloured region around each line. The black lines show how different values of the disc radial scale length to convert the stellar mass to the local stellar surface density affect the  \citet{vanDokkum2013} fit. The dotted line, dashed line and dot-dash line use $R_{scale}$=2.1, 2.4, 1.8 kpc. These values are taken from \citet{Porcel1998}.}
	\label{fig:vanDokkum}
\end{figure}

Fig. \ref{fig:vanDokkum} compares the results of our runs with the stellar mass growth for Milky Way-type galaxies by  \citet{vanDokkum2013}. In this work they fit the stellar mass as a function of redshift for Milky Way type galaxies with a best-fit polynomial,

\begin{equation}
log(M_{star}) = 10.7 - 0.045z - 0.13z^2,
\end{equation}
\noindent
where M$_{star}$ is the stellar mass, and $z$ is the redshift. We convert from stellar mass to the local stellar surface density via,

\begin{equation}
\Sigma_{star} =M_{star}\pi r_{scale}^2 \exp(-r_\odot / r_{scale}) ^2 ,
\end{equation}
\noindent
where r$_\odot$ is the solar radius (8 kpc), and r$_{scale}$ is the disc scale length of 2.4$\pm$0.3 kpc given by \citet{Porcel1998}.

We also include the range of values recovered from the 24 bootstrapped samples of the observations. The stellar mass growth is similar to the \citet{vanDokkum2013} result, except slightly more rapid at early times. This is easily accounted for because the  \citet{vanDokkum2013} result is an average of the growth of many galaxies, with a range of properties. In general, half the stellar mass is in place within 5 Gyr in both the \citet{vanDokkum2013} result and the different runs of the model, which confirms our results from \citet{Snaith2014, Snaith2015}. 

The range of possible values of the radial scale length can strongly impact the growth rate of our model galaxy compared to the \citet{vanDokkum2013} result. \citet{Bland-Hawthorn2016} report a range of radial scale lengths for the disc, ranging from 1.8--3 kpc for the thick and thin discs. Much of this range is captured in Fig. \ref{fig:vanDokkum}. \citet{Sharma2020}, however, assume an evolving radial scale length, while  the analysis of \citet{vanDokkum2013} presents a self-similar evolution for Milky Way-type galaxies. By exploring the evolution of our model's SFH with a range of radial scale lengths we can see the envelope of possible stellar mass growths. The effect of the different scale lengths is illustrative. We do not take into account the effect of a possible evolution of the scale length between the thick and thin discs, because the thin disc scale length depends significantly on the metallicity of the stars on which it is measured, see the analysis of the scale-length of mono-abundance populations by \citet{Bovy2012, Bovy2016c}. These authors  find that the scale lengths can vary from less than 2 to more than 4 kpc depending on their abundance values.

\subsection{Impact of the Guiding Radius}
\label{sec:guiding}

In the previous sections we have concentrated on using the DFs from stars with a current galactocentric radius between 4--6 kpc based on the \texttt{Starhorse} distances. However, such results can be strongly influenced by radial motions, such as blurring. \citet{Leung2019bb} and \citet{Mackereth2018} calculate the orbits of stars in APOGEE using the \texttt{galpy} code \citep{Bovy2015py, Bovy2014py} and the galactic potential derived by \citet{Bovy2015}. This is an axisymmetrical potential and does not contain a bar. This highlights one of the potential issues with using the guiding centre, which requires assumptions about the shape of the Galactic potential, which remains fairly uncertain. Thus, we chose to use the current distances from \texttt{Starhorse} for the majority of this paper, which are similar to the radii calculated from Gaia data. 

\begin{figure} 
	\centering
	\includegraphics[scale=.65]{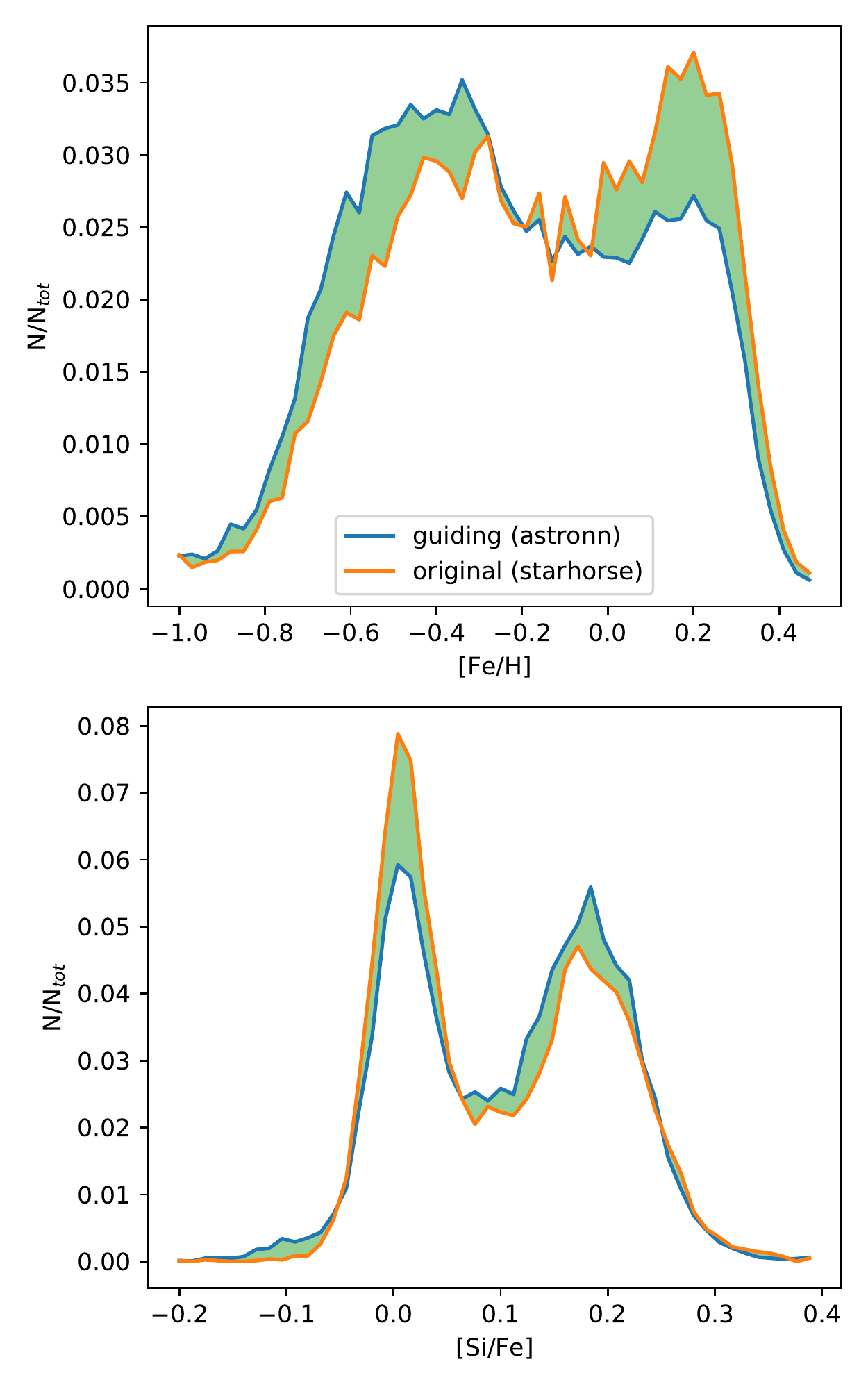}
	\caption{The different distribution functions using the \texttt{Starhorse} and guiding radii. The green area highlights the difference between the two distributions.}
	\label{fig:guidehorse}
\end{figure}

Nevertheless, we explore the consequences of using the guiding radii supplied in the \texttt{astronn} VAC to our result. We found that the $\alpha$DF in particular remains fairly robust, see Fig. \ref{fig:guidehorse}. This figure shows the MDFs (top) and $\alpha$DFs (bottom) of the stars with a guiding or current (\texttt{Starhorse}) galactocentric radius  between 4--6 kpc. The green region emphasises the difference between the DFs defined using the different radius estimations. We note that the MDFs show a significant difference, with a stronger high metallicity peak using \texttt{Starhorse} radii. In fact, for the guiding radii the high metallicity peak is so small it is not readily apparent. On the other hand, there is small increase to the size of highest alpha peak in the $\alpha$DF. These changes to the MDF and $\alpha$DF are consistent, with early star formation being favoured over the later evolution when we make use of the guiding centre radius. 

This can be expected to have an effect on the star formation and infall history derived by our model. The shape of guiding radius DFs should bias the result towards early times, as the low metallicity and high alpha regions of the DFs are more strongly favoured. We also find that the population of stars between 4--6 kpc using the guiding radius is considerably larger. This is because although the APOGEE survey tends to find more stars closer to the Sun due to being magnitude limited, those stars can have guiding radii considerably different from the solar circle.  This means that some of the stars with current positions close to the Sun have guiding radii between 4--6 kpc due to the ellipticity of their orbits. Because the orbits of stars in the thick disc are more elliptical than in the thin disc, this will create a selection effect that will under-represent thin disc stars in the DFs calculated using the guiding radius.

\begin{figure*} 
	\centering
	\includegraphics[scale=.65]{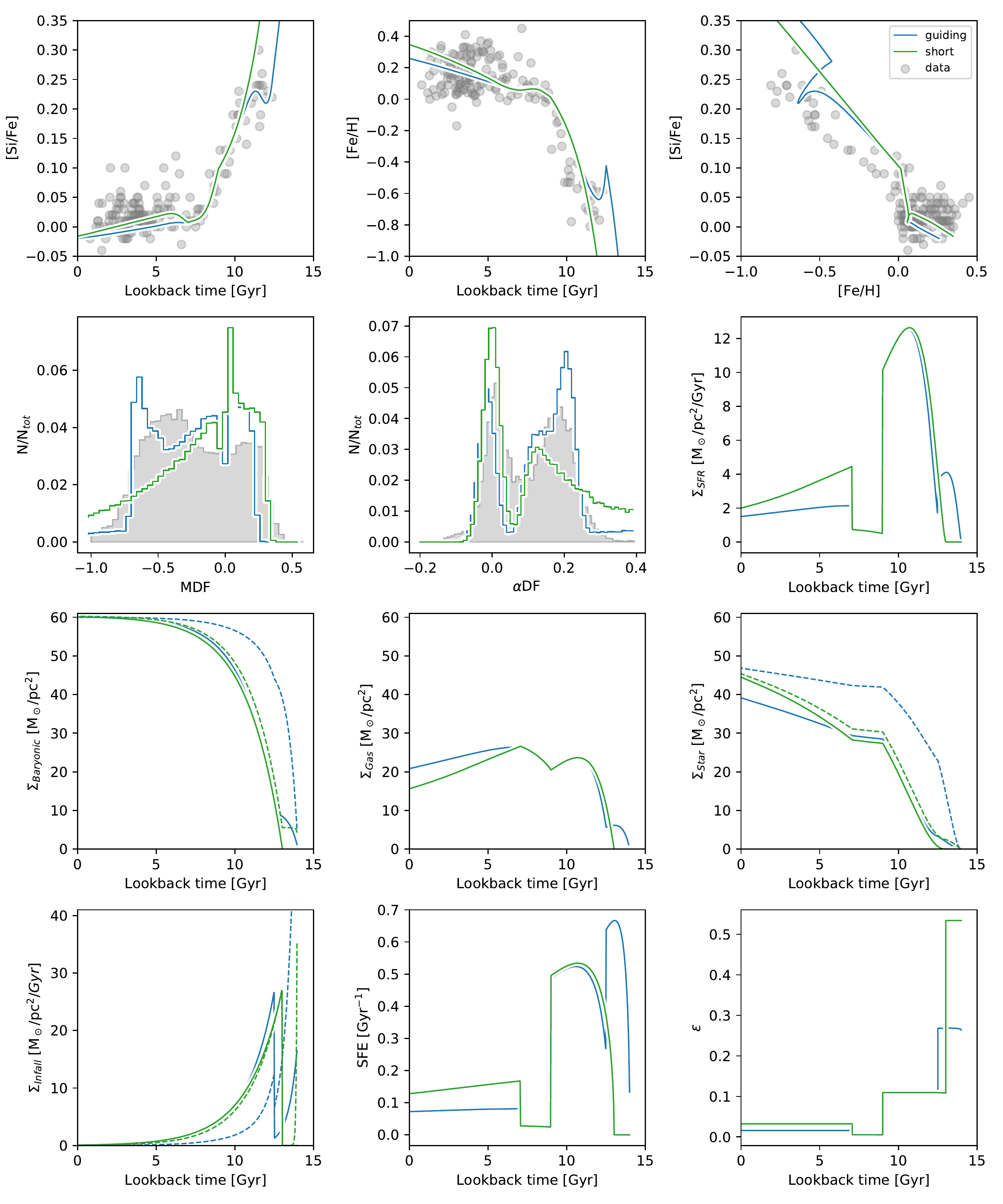}
	\caption{Shows the star formation, infall and chemical evolution of models when assuming  $\omega_{MDF}$=900 and $\omega_{\alpha DF}$=900 where the DFs are defined using the \texttt{Starhorse} radii (green line) and the guiding radii (blue line). Default is the 4--6 kpc bin we have used in the previous sections.}
	\label{fig:guiding}
\end{figure*}

We use the exponential infall model from Section \ref{sec:expinf} to explore how the new DFs affect the mass assembly history. In Fig. \ref{fig:guiding} we show the outputs of the model fitted to the DFs with $\omega_{MDF, \alpha DF}=900$. We then compared this result to the best fit `short' run previously shown in Fig. \ref{fig:compareExpo}.   Using either radius will introduce selection effects, but without careful dynamical modelling beyond the scope of this work it would be difficult to take these effects fully into account. 

The resulting fits are reasonably similar to the previous results. The baryonic formation time is between 2--2.5 Gyr, and the stellar formation time is between 3.5--4 Gyr, which are very close to our the previous runs using the \texttt{Starhorse} radii. We find that the new best-fit result requires a lower star formation history during the thin disc phase, and a stronger initial infall relative to the result for the $\tau_{ini}$=2 Gyr run. This means that the current time stellar surface density has dropped from around 50 M$_{\odot}$/pc$^2$ to 40 M$_{\odot}$/pc$^2$. The baryonic and stellar formation times are largely unchanged at 2.5 Gyr and 3.9 Gyr respectively. 

As previously discussed, the final shape of the star formation and accretion history is relatively insensitive to the chosen initial conditions, so we present the fits we have discovered, which are not necessarily the only solution. There is also a degree of inflexibility to the model, which prevents it from perfectly mimicking the observations. 

One of the limitations of the present study is the absence of radial migration modeling. This is less of an issue when investigating the inner disc (R$<$6 kpc) however, because of the smaller spread in metallicity of the thin disc. At larger radii (R$>$7 kpc), the effect of radial mixing is  the subject of considerable debate, \citep[see, for example ][for some different perspectives on this discussion] {Sharma2020, Halle2015, Khoperskov2021}.

Changing the radii used to define the DFs does not significantly affect the results of the model in terms of the baryonic and stellar formation times, within the model's constraints, and does not affect our conclusions. This suggests that our results are robust to the choice of stellar galactocentric radii chosen to define the distribution functions.

\subsection{The radial evolution -- changing behaviour with radius}
\label{Sec:radiusprof}

We have chosen the region with a current galactocentric (\texttt{Starhorse})  radius between 4--6 kpc for the majority of our studies. We chose this region to avoid the known transition from inner to outer discs around the solar circle ($< 6$ kpc), and to reduce the impact of the bar ($>4$ kpc). Our model, therefore, reproduces the average behaviour of stars in this region. However, the 2 kpc width of this region offers the potential for varying behaviour with radius. In the previous sections we have studied the entire region, in order to gain insight into the average properties of star formation and infall in this region. In this section we will explore the radial effects. 

\begin{figure} 
	\centering
	\includegraphics[scale=.5]{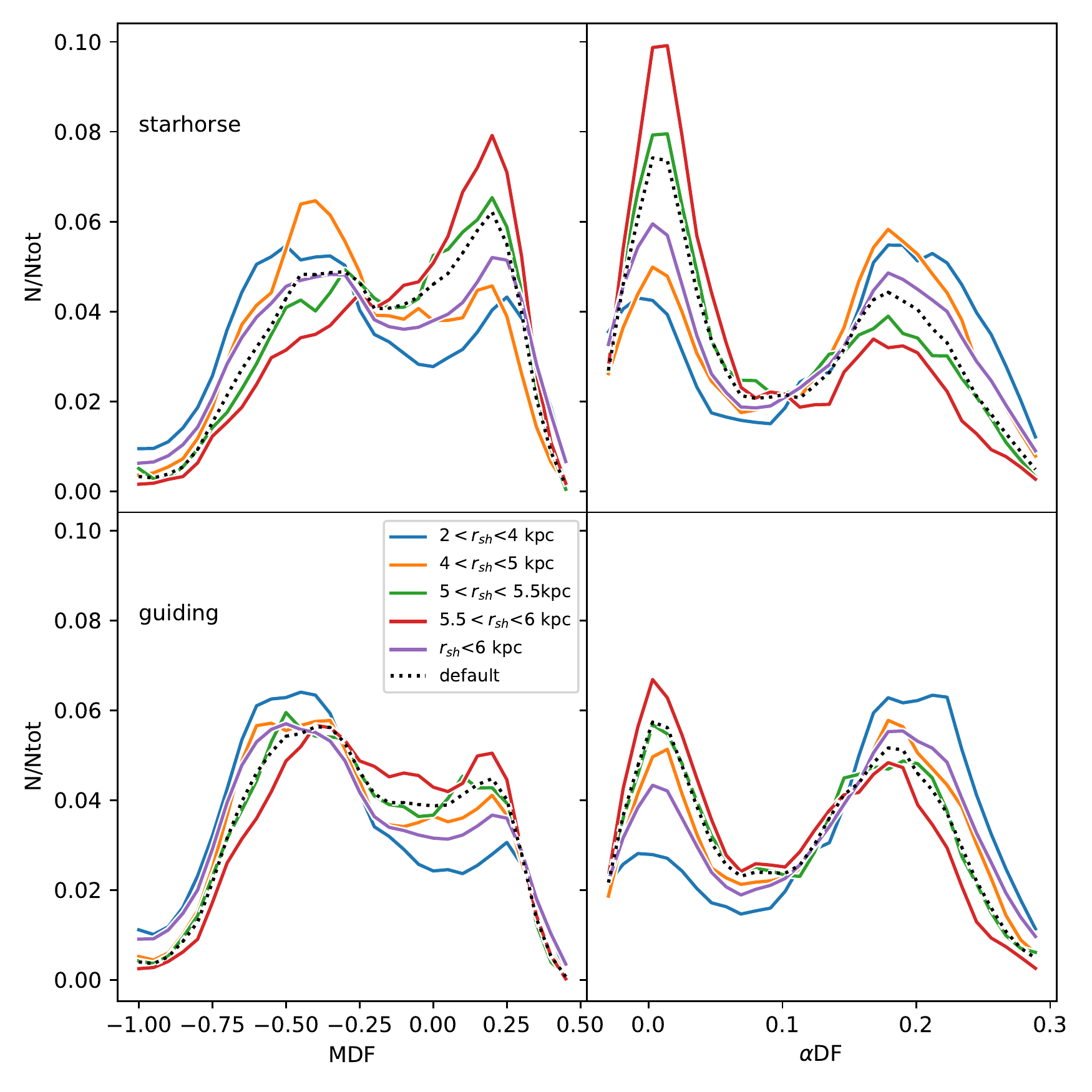}
	\caption{Top row: DFs defined using the \texttt{Starhorse} radii with different radial cuts. Bottom row: DFs defined using the guiding radii  with different radial cuts. Curves are smoothed using a Savitzky-Golay with a window of 5 and a polynomial order of 2.}
	\label{fig:radialbinsobs}
\end{figure}

We show our decomposition of the MDF and $\alpha$DF by radius in Fig. \ref{fig:radialbinsobs}. Although there is variation over the 4--6 kpc region, the overall 4--6 kpc MDF and $\alpha$DF matches well with the smaller 5--5.5 kpc bin. For $r<5$ kpc the DFs favour the low metallicity, high alpha peaks, and $r> 5.5$ very strongly favours the high metallicity peak, without the bimodality seen at smaller radii. The $\alpha$DF retains some bimodality outside of 5 kpc, but the high alpha peak is much weaker. 

We include the MDF and $\alpha$DF of all stars with galacto-centric radii < 6 kpc, which meets our definition of the entire inner disc. The properties of stars within 5 kpc is approximately uniform, but outside this region the behaviour begins to change. Similarly, comparing the MDF between 4--6 kpc and <6 kpc we see only a slight deviation at high metallicity. Although the high metallicity peak is smaller,  there is a good match between the two distributions. This is also true for the $\alpha$DF, where the late time star formation peaks are weaker for the <6 kpc DFs. 

\begin{figure*} 
	\centering
	\includegraphics[scale=.6]{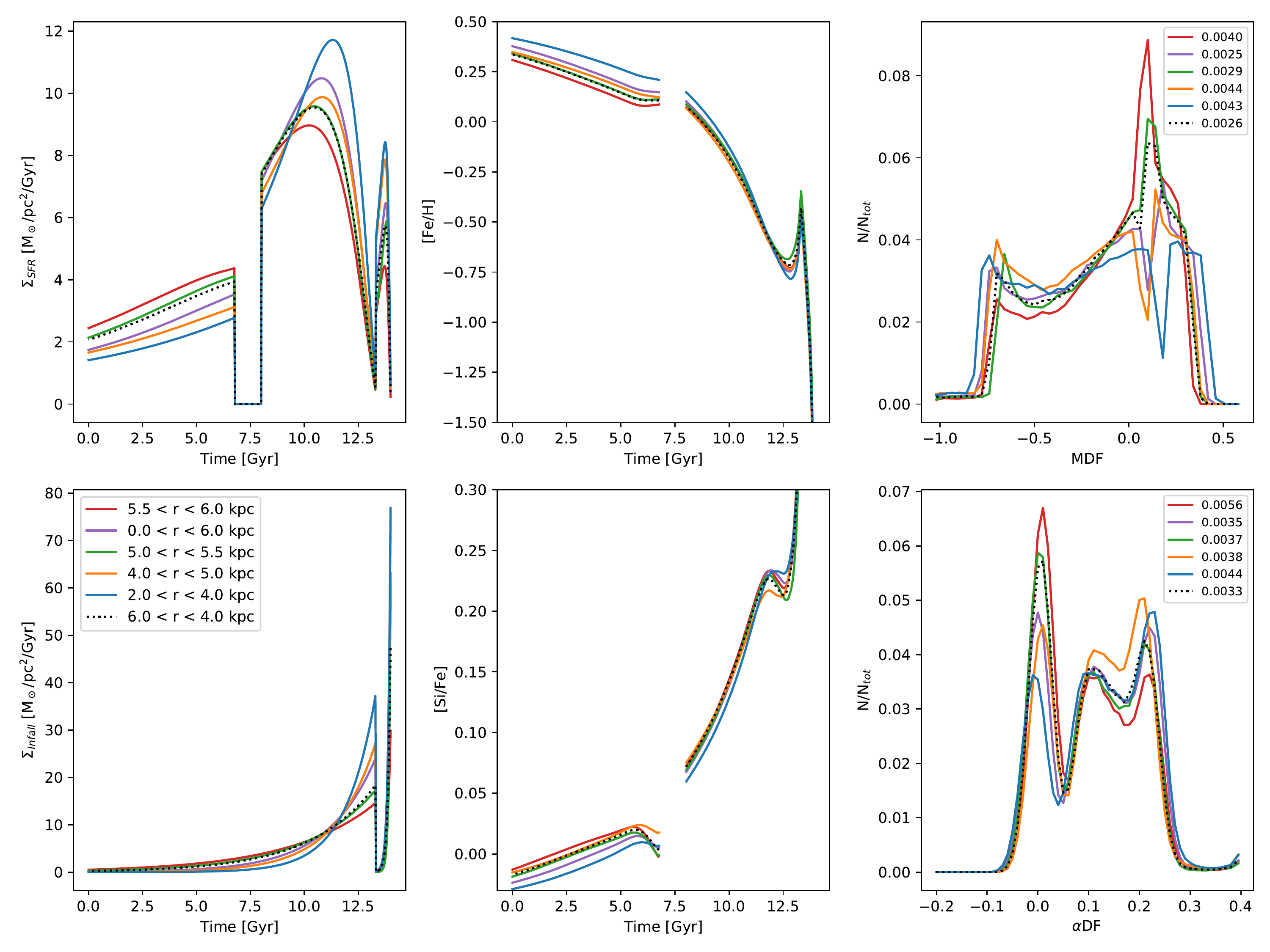}
	\caption{The best fit infall and star formation histories for the different radial cuts (shown in kpc) for the exponential infall model using the initial $\tau$=5. initial conditions. These runs use $\omega_{MDF, \alpha DF}$=900. The different MDFs and $\alpha$DFs in the different radial bins are shown in Fig. \ref{fig:radialbins}. The fitting values between the model and observations for the MDF and $\alpha$DF are shown in the respective panel }
	\label{fig:radialbins}
\end{figure*}	

We fit the model with the DFs shown in Fig. \ref{fig:radialbinsobs}  ($\omega_{MDF, \alpha DF}$=900) using the ICs of the 'medium' value of $\tau$ (from section \ref{sec:expinf}), shown in Fig.  \ref{fig:radialbins}.  We find only a small variation in the recovered infall and star formation histories, although this can have a noticeable effect on the distribution functions. We find that from inside to outside the baryonic formation time is 1.43, 1.71, 2.60, 3.03 Gyr, so there is a noticeable increase in the formation time with (\texttt{Starhorse}) radius. Similarly, the star formation time is increases from 3.75 to 5.02 Gyr. The growth between the 2--4 kpc and 4--5 kpc bins increases the baryonic formation time by 0.3 Gyr, while from 5--5.5 kpc the growth in formation time is 0.9 Gyr and 0.4 Gyr from 5.5 to 6 kpc. The inner regions of the disc show less variation than the outer parts. The fitting parameter for the MDF shows similar values for the 4--6 kpc result (shown previously), the 5--5.5 kpc bin (which we noted is very similar to the to the 4--6 kpc DFs) and the 0--6 kpc cuts, with values of 0.0026, 0.0029 and 0.0025 respectively. Similarly, for the $\alpha$DF the values of the fitting parameter are 0.0033, 0.0037 and 0.0035. The worst fits for both MDFs are for the 2--4 kpc bin which was not included in the original modelling, and the 5.5--6 kpc bin which shows the strong high metallicity/ low alpha peak and a lack of bimodality. These have values for the MDF ($\alpha$DF) of 0.0043 (0.0044) and 0.0040 (0.0056) respectively. 

This suggests that the 0--6 kpc disc can be captured by the model and  that the baryonic and stellar formation times are 2.0 Gyr and 4.3 Gyr respectively. These values are slightly shorter than for the 4--6 kpc bin, but only enhances our conclusion that most gas infall occurred very early. 

The APOGEE sight lines vary with galactocentric radius, due to the geometry of the position of the Sun in the Galaxy. Thus, our conclusions in this section should be interpreted carefully due to potential selection effects. For example, stellar populations are distributed differently in terms of radial and vertical scale length, for different populations, for example see \cite{Bovy2012}, and this may affect the DFs. This is one potential explanation (at least in part) of the changing behaviour of the DFs with radial scale length shown in Fig. \ref{fig:radialbinsobs}.  Nevertheless, our results demonstrate hints of the inside-out formation of the galaxy, where outer regions have larger star formation.   
 
We see that the high metallicity peak in the MDF is much stronger in the outer bins, as expected from the observations. This is due to the higher late-time star formation rate. This suggests that our baryonic and stellar formation times are an upper limit.

\section{Discussion }
\label{Sec:Discuss}

The infall model presented in this paper fits the chemical evolution of the high-$\alpha$ sequence in solar vicinity data very well,  but also the MDF and $\alpha$DF of the inner disc. There is very little difference between how precisely the free infall model can match the data compared to the exponential infall, and in each case the models favour a rapid early infall. Each infall scenario predicts a baryonic formation time of between 1 and 2.5 Gyr.

In the infall model presented here, early infall is very rapid in each of the scenarios we explored. In the 'top hat' runs we see most of the gas falls in before 10 Gyr ago at most, while the exponential infall scenarios converge on a baryon formation time of around 2.5 Gyr. 
This means that in most of the models constrained using the abundance-time, $\alpha$DF and MDF, half the baryon mass of the inner disc was present by 11.5 Gyr ago -- which is a good fit to the closed box approximation used in our previous works \citet{Snaith2014,Snaith2015, Haywood2016}. It shows that assuming an instantaneous accretion of all the gas was a good approximation.

In fact, the closed box model, as shown in \citet{Haywood2016}, appears to reproduce the MDF even more effectively. This is due to a number of reasons. The SFH of the closed box model can be more carefully fitted to the data because we fitted the SFH directly, in approximately 0.5 Gyr bins, while in the infall model we have less direct control over the star formation rate as we only fit the infall and global star formation efficiency. How closely the evolution `reacts' to changes in infall is related to how much gas is present in the system, governed by $\epsilon$.

However, the difficulty the function optimisation code has to fit the star formation efficiency means that there may be other star formation efficiencies and infall parameters that are a closer fit, but are not identified. 

In order to improve this fit we added explicit quenching into our runs, based on lessons learned from \citet{Snaith2014}. This resulted in a better match with observations in runs fitted to the distribution functions. 
We still cannot distinguish between scenarios where the infall has a constant or varying $\epsilon$, but our results suggest that some form of quenching is required at the transition between the thick and thin disc eras  in the 'tophat' runs, although the case for quenching in the exponential model is more ambiguous, see the Appendix for more details.

However, at  the time of quenching there are still considerable quantities of gas present in the system. Thus, there is a need to consider additional physics in order to quench the gas rich galaxy. \citet{Haywood2016} and \citet{Khoperskov2018} identify one possible explanation for this temporary suspension of star formation -- bar quenching. This is a form of morphological quenching, where the formation of the bar can decrease the star formation rate by a factor of ten in a span of less than 1 Gyr, particularly in the inner regions of the disc. This can explain the hiatus in the star formation rate of the galaxy in both the infall and closed box models, even in the presence of significant quantities of gas.  In this scenario the turbulence in the ISM is increased, reducing the star formation efficiency, and is supported by the observations of \citet{Consolandi2016} who find that the inner regions of galaxies with bars are redder than in bar-less galaxies.  Alternatively, a galaxy-galaxy interaction could have caused the quenching era, as it seems to coincide with the end of the thick disc and a major accretion event \citep[e.g.][]{DiMatteo2019}. 

 In the inner regions of the disc,\ the bar can be expected to shape the star formation history in ways not captured by our model. It  is possible that the ability of the bar  to suppress star formation could mean that in the inner regions of the Milky Way the star formation is prolonged relative to a simple Schmidt-Kennicutt law. This would allow the baryonic accretion of the inner regions to be even faster than estimated by our model \citep{Khoperskov2018}. It is also possible that the radial movements of the gas can redistribute matter prior to star formation, changing the infall history of the Milky Way in ways our model cannot capture. This is beyond the scope of this paper, but would provide for future research using hydrodynamical simulations.   

To our knowledge, there is no equivalent in the literature to the model proposed here (with the exception of the closed-box model in \citet{Haywood2016}), because, traditionally, studies of  chemical evolution model have not been designed to represent only the inner disc evolution.  The only region in the Galaxy where the MDF can be said to be representative of the high-alpha sequence is in the inner disc (R$<$6~kpc). 

 High alpha sequence stars are a minority at the solar vicinity, with the sample
being dominated by solar metallicity stars that {\it are not} on the high-alpha sequence.
The model of \citet{Spitoni2019} is  a recent representative of the class of models that tries to match the entire solar vicinity distribution. In their case, the amount of first infall episode is fixed according to the density of thick disk in the solar vicinity, which we believe is not representative of the thick disc as a whole. Hence, their result that the fraction of the stellar mass in the thick disc stars is just above 10$\%$ of the total surface density, while, in our model it is around 50$\%$. As mentioned in \citet{Haywood2019}, while the age-chemistry relations of this population is probably an effect of the overall mixing that occurred at this time in a population of a few 
10$^{10}$~$\rm M_{\odot}$, the number density of this population in the solar vicinity is only a small percentage of the overall fraction of this population. Specifically, half of the stellar mass of the inner disc is in the thick disc component, formed more than 8 Gyr ago, while at the solar circle the thick disc stellar mass fraction is only 10\%, because of the different radial scale lengths of the two components \citep{Bovy2012}.

Despite this, in our previous work \citep{Snaith2014, Snaith2015} we were able to make use of data in the solar vicinity to model the inner disc. In this case, however, we first separated the high and low alpha sequences \citep[as in][]{Haywood2013} and fitted them separately. This is equivalent to our current decomposition based on distance from the galactic centre. In addition, in our previous work we did not fit the DFs, only the region of age-[Si/Fe]-[Fe/H] space populated by those stars. It was in  \citet{Haywood2016} that the strong correspondence between the DFs produced by our model and the DFs of the inner disc was noted.

Although we pointed out that our fits to the data produce a stellar and gas surface density comparable with observations of the solar vicinity, this is not necessarily relevant, as the high alpha sequence is expected to correspond to the entire inner disc. This means that global measurements from the solar vicinity are not particularly helpful. The model should be compared to the entire inner disc. Although we fixed the total baryon fraction to be 60 M$_\odot$/pc$^2$ this is not required, for these same reasons. We scale $\epsilon$ by $\sqrt(\Sigma_{bary})$, and so the evolution is effectively scale free.

\section{Conclusions}
\label{Sec:Conc}
We have elaborated on our previous work \citep{Snaith2014,Snaith2015} to include explicit rather than implicit infall of gas in our chemical evolution model. We have fitted our models to updated stellar ages based off \citet{Adibekian2012} and Gaia, and made use of the MDF  and $\alpha$DF from APOGEE DR16 to refine our best-fit star formation histories. We expanded the closed box model of the inner galactic disc to include explicit infall and the Schmidt-Kennicutt law between gas surface density and star formation rate density. 

Our free infall, or 'top hat' infall model uses fourteen 1 Gyr infall episodes to model the infall history of the Milky Way, whilst our exponential infall model includes four eras of star formation, the initial infall, the thick disc, the quenching and the thin disc. The quenching epoch is inserted by hand in both sets of scenarios, and has a low (or zero) star formation efficiency. 

Our main conclusions are summarised as:

\begin{itemize}
       \item The baryon formation time of the inner disc of the Milky Way was rapid, taking around 2 Gyr to assemble half its present day baryonic (star + gas) mass. The stellar formation time, defined as the time taken for half the mass of stars to form, is longer, taking around 5 Gyr. This corresponds well with our prediction from \citet{Snaith2014} that the thick disc contains half the stellar mass of the inner regions of the Milky Way, and that the inner disc can be approximated with a closed box. 

	\item The MDF and $\alpha$DF are essential for breaking the degeneracy of the different infall parameters, but the time-abundance distribution is also needed to capture the important features of the evolution.
	 \item The exact choice of galactocentric radius used to define the APOGEE subsample used to build the distribution functions can affect its overall shape. However, within the limitations of the model, this does not strongly impact our results.
	\item  Similarly, the selection function of APOGEE well represents the inner Galaxy, and does not strongly affect the results of our models.
	\item The best-fit infall parameters we have used produce a very similar chemical  evolution track to the observations, because this infall model has most of the gas in place early, and is therefore similar to the closed box model overall. 
	\item A quenching epoch is  used for recovering the key features in the chemical evolution and MDF in both the infall and closed box models, as first proposed in \cite{Haywood2016}. Without quenching, the characteristic sharp change in gradient in the age-[Si/Fe] distribution is not recovered by the 'tophat' infall model, and the MDF and $\alpha$DF do not show the twin peaked distribution seen in APOGEE.  We discuss the quenching in the exponential infall model in more detail in the Appendix. The precise shape of the dip is difficult to capture, even in runs with $\omega_{MDF,\alpha DF}$=900, because of the need to fit both the $\alpha$DF and MDF at the same time.
	\item A possible mechanism responsible for the star formation hiatus is bar quenching, which is different to the method proposed by \citet{Noguchi2018}. This is because there is considerable gas present in the ISM when the quenching is required, so a lack of gas cannot be used to change the gradient of the [Si/Fe] evolution in the way we see in the observations. Alternatively, the end of satellite accretion \citep[e.g.][]{Kruijssen2019} could have resulted in a change of conditions in the disc and led to quenching, or the emergence of the bar and the resulting redistribution of gas could have suppressed star formation. 
	\item All the models reproduce the massive thick disc we proposed in \citet{Snaith2014,Snaith2015}  for the inner regions of the Milky Way. 
	\item Additional physics beyond the Schimdt-Kennicutt star formation law is needed to fully understand the star formation history of the Milky Way. 
	\item Our model is fully compatible with the infall paradigm for galaxy formation with a low star formation efficiency. This low efficiency can be assumed to be due to the gas reservoir into which we mix the metals being larger than the star forming ISM.
	\item Our star formation histories compare favourably with the result of \citet{vanDokkum2013}, implying that the Milky Way is representative of galaxies of its type. 
\end{itemize}

We can characterise the closed box model as an infall model with a low star formation efficiency and a rapid gas accretion rate at early times. In future we will expand our model to account for the outer disc. However, in order to model the outer disc we will need to make use of a multi-zone model, because, as proposed  by \citet{Haywood2013}, we believe that the outer disc initial condition is set by outflows from the inner disc and dilution by low metallicity gas \citep{Snaith2015, Haywood2018}. This has recently been examined by \citet{Katz2021} using APOGEE DR16 data, but would benefit from modelling for further insight. Even in the infall model the key components identified in \citet{Snaith2015} remain important, the high SFR in the thick disc, the low SFR in the thin disc and (potentially) the quenching epoch are all important for shaping the chemical properties of the inner galaxy.

\begin{acknowledgements}
The authors thank the referee for their suggestions which enabled us to refine our arguments and our comparison with the observations. We thank Laura Boyle for identifying our error in quoting the equation of \citet{Raiteri1996} in \citet{Snaith2015}.
ONS thanks the DIM ACAV+ for the funding that made this work possible. 
PDM and MH thank the ANR (Agence Nationale de la Recherche) for its financial support through the MOD4Gaia project (ANR-15- CE31-0007, P.I.: P. Di Matteo).
SK acknowledges the Russian Science Foundation (RSCF) grant 19-72-20089 for support in the preparation of the N-body models partially carried by using the equipment of the shared research facilities of HPC computing resources at Lomonosov Moscow State University (project RFMEFI62117X0011).
This work was granted access to the HPC resources of CINES and TGCC-CEA under the allocation 2018-0410154 made by GENCI.
SK acknowledges the Russian Science Foundation (RSCF) grant 19-72-20089 for support in  the preparation of the N-body models.
	Funding for the Sloan Digital Sky Survey IV has been provided by the Alfred P. Sloan Foundation, the U.S. Department of Energy Office of Science, and the Participating Institutions. SDSS-IV acknowledges
	support and resources from the Center for High-Performance Computing at
	the University of Utah. The SDSS web site is www.sdss.org.
	
	SDSS-IV is managed by the Astrophysical Research Consortium for the 
	Participating Institutions of the SDSS Collaboration including the 
	Brazilian Participation Group, the Carnegie Institution for Science, 
	Carnegie Mellon University, the Chilean Participation Group, the French Participation Group, Harvard-Smithsonian Center for Astrophysics, 
	Instituto de Astrof\'isica de Canarias, The Johns Hopkins University, Kavli Institute for the Physics and Mathematics of the Universe (IPMU) / 
	University of Tokyo, the Korean Participation Group, Lawrence Berkeley National Laboratory, 
	Leibniz Institut f\"ur Astrophysik Potsdam (AIP),  
	Max-Planck-Institut f\"ur Astronomie (MPIA Heidelberg), 
	Max-Planck-Institut f\"ur Astrophysik (MPA Garching), 
	Max-Planck-Institut f\"ur Extraterrestrische Physik (MPE), 
	National Astronomical Observatories of China, New Mexico State University, 
	New York University, University of Notre Dame, 
	Observat\'ario Nacional / MCTI, The Ohio State University, 
	Pennsylvania State University, Shanghai Astronomical Observatory, 
	United Kingdom Participation Group,
	Universidad Nacional Aut\'onoma de M\'exico, University of Arizona, 
	University of Colorado Boulder, University of Oxford, University of Portsmouth, 
	University of Utah, University of Virginia, University of Washington, University of Wisconsin, 
	Vanderbilt University, and Yale University.
	This work has made use of data from the European Space Agency (ESA)
mission {\it Gaia} (\url{https://www.cosmos.esa.int/gaia}), processed by
the {\it Gaia} Data Processing and Analysis Consortium (DPAC,
\url{https://www.cosmos.esa.int/web/gaia/dpac/consortium}). Funding
for the DPAC has been provided by national institutions, in particular
the institutions participating in the {\it Gaia} Multilateral Agreement. 
We have made use of \texttt{numpy} \citep{2020NumPy-Array},  \texttt{scipy} \citep{2020SciPy-NMeth},  \texttt{matplotlib} \citep{4160265},  F2PY \citep{peterson2009f2py} and  \texttt{pandas} \citep{reback2020pandas} libraries in the  \texttt{Python} language as well as \texttt{IPython} \citep{4160251} and \texttt{Jupyter notebook} \citep{Kluyver2016a}.

\end{acknowledgements}

\begin{appendix}
	\section{Quenching in the Exponential Infall Model}

\begin{figure} 
      		\centering
      		\includegraphics[scale=.5]{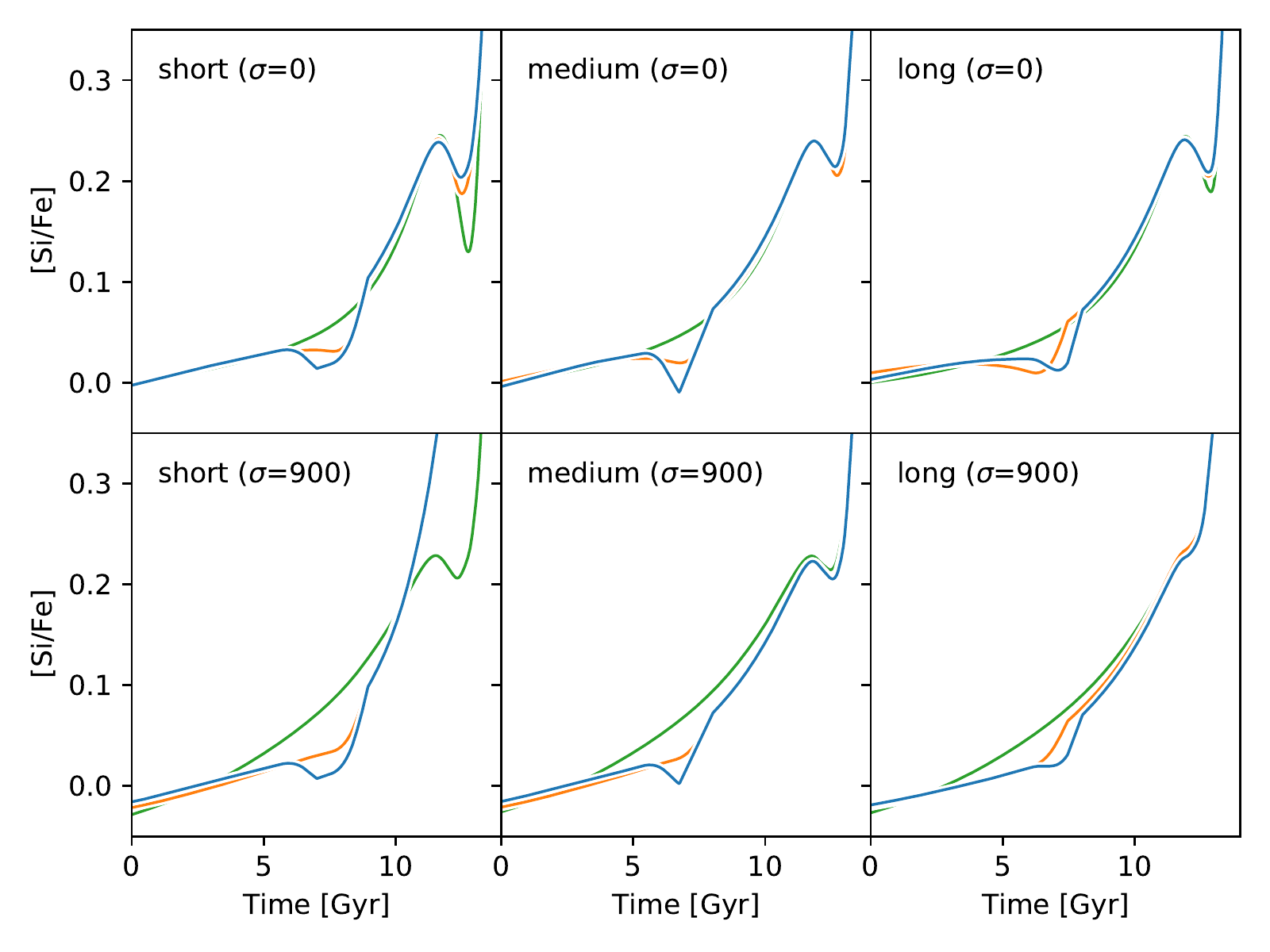}
      		\caption{The different star formation histories fitted using different initial star formation histories. The blue line shows the [Si/Fe]-time evolutions from section 4.1. The orange lines show the result [Si/Fe]-time evolution with the explicit quenching removed, and the green line shows the result when the late time thin disc $\epsilon$ is set to be the same as the thick disc $\epsilon$.}
      		\label{fig:comparequenchsife}
      	\end{figure}

\begin{figure} 
      	\centering
      	\includegraphics[scale=.5]{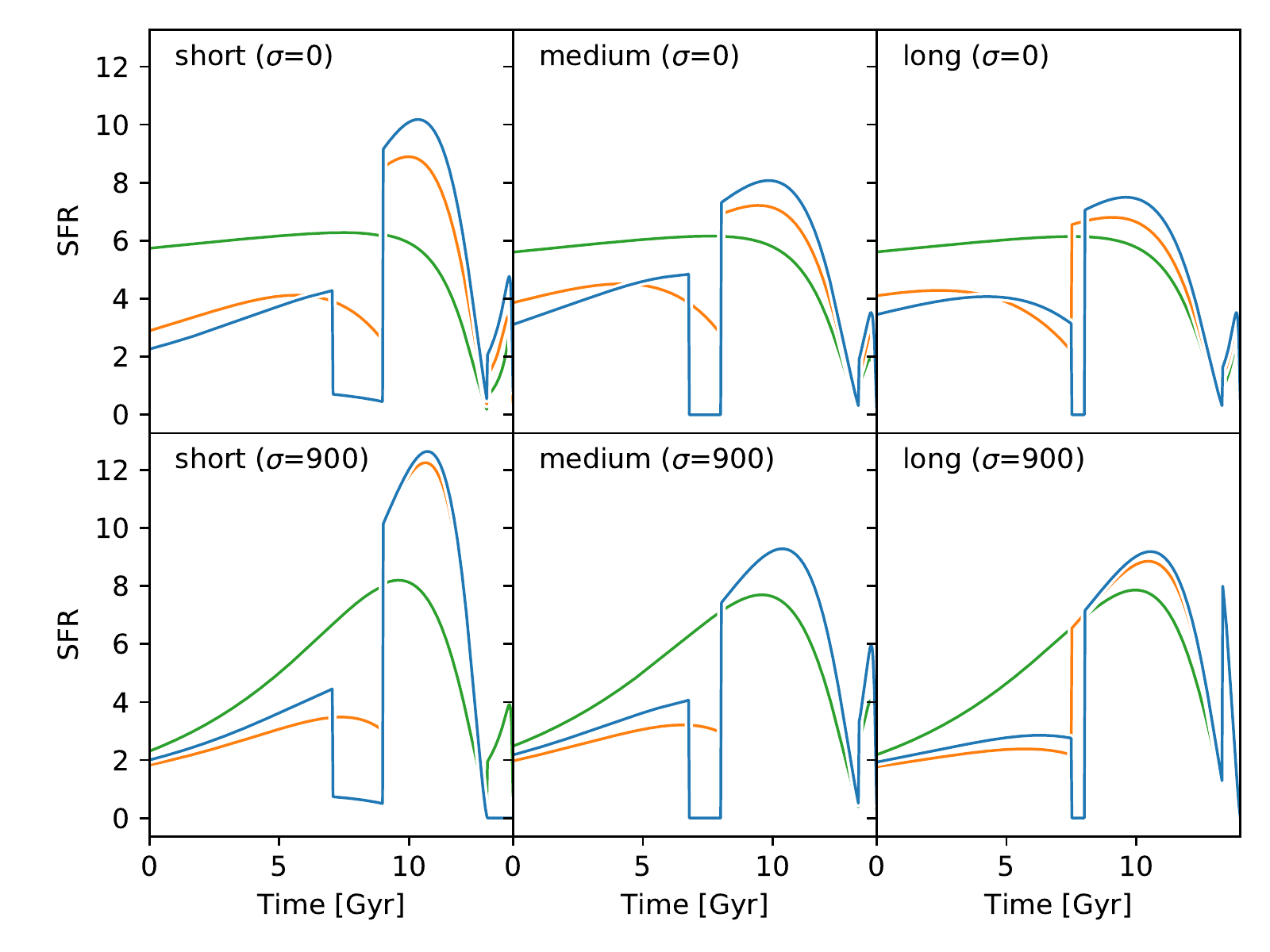}
      	\caption{The fitted star formation histories related to the chemical evolutions presented in Figure \ref{fig:comparequenchsife}.}
      	\label{fig:comparequenchsfh}
      \end{figure}

 	We included an explicit quenching episode in our initial star formation histories for the exponential infall scenario. This was driven by our result from \citet{Haywood2016}, and  \citet{Snaith2014, Snaith2015}. However, in Fig. \ref{fig:comparequenchsife} we see that the sharp transition between the thick and thin disc eras can be formed  with a sharp change in $\epsilon$ rather than explicit quenching (the corresponding star formation histories are shown in Fig. \ref{fig:comparequenchsfh}). 
	
	In order to fit the data using the exponential infall model it is not required that we include the quenching era where $\epsilon$ drops to zero, or close to zero, and is followed by a reignition of star formation. A sharp transition from high to low $\epsilon$ of around a factor of 3 is, however, required to reproduce the sharp gradient change in the [Si/Fe] evolution. In Fig. \ref{fig:comparequenchsife} the blue and orange lines are essentially indistinguishable. If we bootstrapped the data and made the fit multiple times, any different would be lost in the error.  This result is replicated in the DFs, where the explicit quenching run (blue line), and the $\epsilon$ reduction run (orange line) are not significantly different. On the other hand, where there is no strong change in the star formation efficiency at around 8 Gyr ago, there is no sharp transition in the [Si/Fe]-time evolution, and the DFs do not reproduce the data as well. 
	
	At first sight, this result would appear to be in tension with our previous work in \citet{Snaith2014, Snaith2015} which required an explicit quenching episode. However, the closed box model is able to much more effectively shape the star formation history to the data, and so more carefully match the chemical evolution of the system. This is because the exponential infall model is more highly constrained than the closed box model, as it assumes: (1) that infall follows an exponential form, (2) the star formation rate is affected only by the amount of gas present and $\epsilon$, (3) there are only three eras in which the infall parameters, and $epsilon$ can change and (4) we control the amount of gas in the system not the star formation rate directly. In the closed box model there were twenty eight eras, each of which could change to shape the chemical evolution track. Thus, the closed box model is able to more precisely model the chemical evolution of the stars in our sample. Both models (exponential infall and closed box) show that the data outside the transition region can be matched by a star formation rate which is around 3 times lower in the thin disc phase than in the thick disc phase. 
	
	In \cite{Haywood2016} the dearth of intermediate values in the $\alpha$DF is used to argue for a quenching era in the Milky Way, (more specifically the scenario where we have high star formation followed by very low star formation for 1 Gyr and a re-ignition of star formation at a lower value). Our result here does not exclude this, because, as discussed, the closed box model  better fits the [Si/Fe]-age data, and the DFs, compared to the exponential infall model. In Fig. \ref{fig:compareExpo} and \ref{fig:compareExpo900} we can see that although the exponential infall model correctly reproduces the key features of the $\alpha$DF (the two peaks and the dip) the dip is not as broad. Such a broad dip was better reproduced using our closed box model, as shown in Fig. 4 of  \cite{Haywood2016}.

\end{appendix}

\bibliographystyle{a_and_a/aa}
\bibliography{bibliography}

\end{document}